\documentclass[11pt]{article}
\usepackage{fullpage}
\usepackage{cite}
\usepackage{graphicx}
\usepackage{amsmath}
\usepackage{amssymb}
\usepackage{subcaption}
\usepackage{mathtools}
\usepackage{xcolor}
\usepackage[most]{tcolorbox}
\usepackage{blindtext}
\usepackage{tikz-cd}
\usepackage{subfiles}
\usepackage{titlesec}
\usepackage{parskip}
\title{}
\date{}

\def\beq{\begin{equation}}
\def\eeq{\end{equation}}
\allowdisplaybreaks
\begin{document}
\bibliographystyle{utphys}
\newcommand{\sech}{sech}
\newcommand{\csch}{csch}

\newcommand{\Ft}{\tilde{F}}
\newcommand{\msbar}{\ensuremath{\overline{\text{MS}}}}
\newcommand{\DIS}{\ensuremath{\text{DIS}}}
\newcommand{\abar}{\ensuremath{\bar{\alpha}_S}}
\newcommand{\bb}{\ensuremath{\bar{\beta}_0}}
\newcommand{\rc}{\ensuremath{r_{\text{cut}}}}
\newcommand{\Nd}{\ensuremath{N_{\text{d.o.f.}}}}
\setlength{\parindent}{0pt}

\titlepage
\begin{flushright}
QMUL-PH-19-32
\end{flushright}

\vspace*{0.5cm}

\begin{center}
{\bf \Large S-duality and the Double Copy}

\vspace*{1cm} \textsc{Rashid Alawadhi\footnote{r.alawadhi@qmul.ac.uk},
David S. Berman\footnote{d.s.berman@qmul.ac.uk},
Bill Spence\footnote{w.j.spence@qmul.ac.uk}  David Peinador Veiga\footnote{d.peinadorveiga@qmul.ac.uk}   } \\

\vspace*{0.5cm} $^a$ Centre for Research in String Theory, School of
Physics and Astronomy, \\
Queen Mary University of London, 327 Mile End
Road, London E1 4NS, UK\\

\end{center}

\vspace*{0.5cm}

\begin{abstract}
The double copy formalism provides an intriguing connection between gauge theories and gravity.
It was first demonstrated in the perturbative context of scattering amplitudes but recently the formalism has been applied to exact classical solutions in gauge theories such as the monopole and instanton.
In this paper we will investigate how duality symmetries in the gauge theory double copy to gravity and relate these to solution generating transformations and the action of $SL(2,\mathbb{R})$ in general relativity.

\end{abstract}

\vspace*{0.5cm}

\section{Introduction}
\label{sec:intro}

The double copy was first investigated in a series of works~\cite{Bern:2008qj,Bern:2010ue,Bern:2010yg} as a relationship between perturbative scattering
amplitudes in gauge theory and gravity. It has been proved at tree
level~\cite{BjerrumBohr:2009rd,Stieberger:2009hq,Bern:2010yg,BjerrumBohr:2010zs,Feng:2010my,Tye:2010dd,Mafra:2011kj,Monteiro:2011pc,BjerrumBohr:2012mg},
where it has a stringy origin~\cite{Kawai:1985xq}. But there is still no non-perturbative proof of the double copy, although the evidence is mounting with a series of papers showing double copy behaviour for amplitudes at higher loop order \cite{Bern:2010ue,Bern:1998ug,Green:1982sw,Bern:1997nh,Carrasco:2011mn,Carrasco:2012ca,Mafra:2012kh,Boels:2013bi,Bjerrum-Bohr:2013iza,Bern:2013yya,Bern:2013qca,Nohle:2013bfa,
  Bern:2013uka,Naculich:2013xa,Du:2014uua,Mafra:2014gja,Bern:2014sna,
  Mafra:2015mja,He:2015wgf,Bern:2015ooa,
  Mogull:2015adi,Chiodaroli:2015rdg,Bern:2017ucb,Johansson:2015oia,Oxburgh:2012zr,White:2011yy,Melville:2013qca,Luna:2016idw,Saotome:2012vy,Vera:2012ds,Johansson:2013nsa,Johansson:2013aca}.

More recently, the double copy/single copy was applied to a class of exact classical solutions. {\it{Double copy}} refers to moving from gauge theory to gravity while {\it{single copy}} is the inverse map from gravity to gauge theory. The Schwarzschild solution was shown to single copy to an electric charge \cite{Monteiro:2014cda} and the Taub NUT solution single copy to a magnetic monopole \cite{Luna:2015paa}. Subsequent to that the single copy of the Eguchi-Hanson solution has been mapped to a self-dual gauge field \cite{Berman:2018hwd}. More general topologically non-trivial solutions have been double copied in the work of \cite{Sabharwal:2019ngs}. (Other work examining symmetries of the linearised double copy have also appeared in a series of works \cite{Anastasiou:2014qba,Borsten:2015pla,Anastasiou:2016csv,Anastasiou:2017nsz,Cardoso:2016ngt,Borsten:2017jpt,Anastasiou:2017taf,Anastasiou:2018rdx,LopesCardoso:2018xes}.)

To futrher investigate the double copy formalism beyond the perturbative regime, we examine how non-perturbative symmetries in the gauge theory are double copied to gravity. In particular, gauge theories exhibit electromagnetic duality which exchanges electric and magnetic charges. This symmetry, often also called S-duality, was first discovered for classical Maxwell theory but then became a crucial ingredient in the study of properties of Yang-Mills theory \cite{Montonen:1977sn}. General relativity is not known to exhibit S-duality which leads to the question of what S-duality could double copy to. We will address this at the non-perturbative classical level by identifying the solution generating symmetry in general relativity that single copies to electromagnetic duality.

We will show that the double copy of electromagnetic duality is identified as the Ehlers transformation in general relativity. It is worth commenting further that we are working with exact and thus classically non-perturbative solutions in the double copy and the metrics corresponding to the {\it{electric}} and {\it{magnetic}} solutions are mapped between each other using an exact non-local transformation. We will then demonstrate how another solution generating symmetry discovered by Buchdahl \cite{PhysRev.115.1325}, is single copied to the charge conjugation.

We adopt two complementary approaches. First we will use the Kerr-Schild form of the metric where the metric in this form is related to gauge fields in the single copy. Second we will develop the correspondence introduced in \cite{Luna:2018dpt} where the Weyl curvature in gravity is related to a combination of field strengths in the single copy. In each case we will examine how the electromagnetic transformation in the single copy is related to a transform in relativity.

To make the paper as self contained as possible and make clear our conventions for the double/single copy we begin with a description of the Kerr-Schild form in GR and then the single copy prescription.{\footnote{Other conventions and methods for the double copy are also available, see  for example \cite{Anastasiou:2018rdx,LopesCardoso:2018xes}.}} We give the detailed examples of the Schwarzschild black hole and the Taub-NUT solution and their single copies as described first in \cite{Monteiro:2014cda,Luna:2015paa}. Then we describe the Ehlers transform in general before moving on to its application to Schwarzchild and its single copy. After discussing the role of the the Buchdahl transform and its single copy again using Kerr-Schild form we move to the double copy in terms of curvatures and field strengths, known as the {\it{Weyl double copy}}  \cite{Luna:2018dpt} where we can again examine the role of electromagnetic duality from a complementary approach. We will show  in a number of cases that the duality transformations in electromagnetism  correspond with solution-generating transformations in general relativity, and together preserve the form of the double copy.

{\bf{Note added}}: While this manuscript was in preparation \cite{Huang:2019cja} appeared which has overlap with this paper.

\section{Classical Double Copy}\label{chap:DC}
\subsection{The Kerr-Schild double copy}
First let use introduce the Kerr-Schild form of the metric and examine the behaviour of Einstein's field equations. Writing  the metric in this form is a crucial step in making the double/single copy  manifest. In what follows we shall use the procedure outlined in \cite{Monteiro:2014cda}. We take the metric $\eta_{\mu\nu}={\rm diag}(-1,+1,+1,+1)$ throughout.

A solution is ``Kerr-Schild"  if a set of coordinates may be found such that the spacetime metric $g_{\mu\nu}$ may be put in the form
\begin{equation}
  g_{\mu\nu}=\eta_{\mu\nu}+\phi k_\mu k_{\nu}\, ,
\end{equation}
where $\phi$ is a scalar field and $k_{\mu}$ is a co-vector satisfying
\begin{equation}
  \eta^{\mu\nu}k_{\mu}k_{\nu}=0=g^{\mu\nu}k_{\mu}k_{\nu}\, ,
\end{equation}
{\it i.e.} it is null with respect to both the full and background metric. The inverse metric then takes the form
\begin{equation}
  g^{\mu\nu}=\eta^{\mu\nu}-\phi k^\mu k^{\nu} \, .
\end{equation}
 In terms of the scalar field $\phi$ and co-vector $k_\mu$, the Ricci tensor and Ricci scalar are
\begin{equation}
  \begin{split}
    &R^\mu_\nu=\frac{1}{2}(\partial^\mu\partial_\alpha(\phi k^\alpha k_\nu)+\partial_\nu\partial^\alpha(\phi k_\alpha k^\mu)-\partial^2(\phi k^\mu k_\nu)) \, ,\\
    &R=\partial_\mu\partial_\nu(\phi k^\mu k^\nu)\, ,
\end{split}
\end{equation}
where $\partial^\mu=\eta^{\mu\nu}\partial_\nu$. In the stationary spacetime case ($\partial_0 \phi=\partial_0 k^\mu=0$) one may take the time component of the Kerr-Schild vector as $k^0=1$, with the dynamics in the time component contained in $\phi$. As a consequence, the components of the Ricci tensor simplify to
\begin{equation}
  R^{0}_{0}=\frac{1}{2}\nabla^{2}\phi\, ,
\end{equation}
\begin{equation}
  R^{i}_{0}=-\frac{1}{2}\partial_{j}[\partial^{i}(\phi k^j)-\partial^j(\phi k^i)]\, ,
\end{equation}
\begin{equation}
  R^{i}_{j}=\frac{1}{2}\partial_{l}[\partial^{i}(\phi k^{l}k_{j})+\partial_{j}(\phi k^{l}k^{i})-\partial^{l}(\phi k^{i}k_{j})]\, ,
\end{equation}
\begin{equation}
  R=\partial_{i}\partial_{j}(\phi k^{i}k^{j})\, ,
\end{equation}
where Latin indices indicate the spatial components.

Now  define a gauge field $A_{\mu}=\phi k_\mu$, with the Maxwell field strength $F_{\mu\nu}=\partial_{\mu} A_{\nu}-\partial_{\mu}A_\nu$. Taking the stationary case of the vacuum Einstein equations $R_{\mu\nu}=0$, one finds that the gauge field satisfies the Abelian Maxwell equations
\begin{equation}
  \partial_{\mu}F^{\mu\nu}=\partial_{\mu}(\partial^{\mu}(\phi k^{\nu})-\partial^{\nu}(\phi k^{\mu}))=0 \, .
\end{equation}

The remarkable thing about the double copy is that if now one considers a non-Abelian gauge field $A^a_{\mu}$ with the gauge group index $a$, there is still a single copy/double copy relationship. The recipe is to take the quantity $\phi k_\mu k_\nu$ of a given gravity solution and strip off one of the Kerr-Schild vectors and dress with a gauge group index to get the corresponding gauge field $A^a_\mu=c^a \phi k_\mu$. There is no derivation as such for this procedure but by now there is a compelling amount of evidence as listed in the introduction.
Thus, the basic statement of the double/single copy we will be applying is:
\begin{tcolorbox}[enhanced jigsaw,
    sharp corners,
    boxrule=0.5pt,
    colback=white!30!white,
    borderline={0.5pt}{-2pt}{black,solid} % 0.5pt linewith, -2pt outside, black solid linestyle
 ]
If $g_{\mu\nu=}\eta_{\mu\nu}+\phi k_\mu k_{\nu}$ is a stationary solution of Einstein's equations, then $A_{\mu}^a=c^a\phi k_\mu$ is a solution of the Yang-Mills equations which is linearised by the Kerr-Schild coordinates allowing an arbitrary choice for the constant $c^a$.
\end{tcolorbox}

\subsection{Schwarzschild and NUT spacetimes}

We shall demonstrate the double/single procedure on two important gravity solutions, the Schwarzschild black hole and the Taub NUT spacetime.

%%%%%%
\subsubsection{Schwarzschild spacetime}
The Schwarzschild solution is
\begin{equation}
  ds^2=-\Big(1-\frac{2GM}{r}\Big)dt^2+\frac{dr^2}{1-\frac{2GM}{r}}+r^2d\Omega^2\, ,
\end{equation}
where $d\Omega^2=d\theta^2+\sin^2\! \theta\, d\phi^2$ is the line element on the unit sphere, $G$ is Newton's constant, and $r^2=x^2+y^2+z^2$ is the radial distance from the origin.

In order to apply the single copy procedure, one must write this metric in Kerr-Schild form. To do this, we apply the coordinate transformation $l=t+\bar{r}$ with $d\bar{r}=(1-\frac{2GM}{r})^{-1}dr$, so that the metric takes the form
\begin{equation}
  ds^2=-dl^2+2dldr+r^2d\Omega^2+\frac{2GM}{r}dl^2 \, .
\end{equation}
A further coordinate transformation of the form $l=\bar{t}+r$ is then applied so that the metric becomes
\begin{equation}
  ds^2=-d\bar{t}^2+dr^2+r^2d\Omega^2+\frac{2GM}{r}(d\bar{t}^2+dr^2+2d\bar{t}dr) \, .
\end{equation}
Notice that the first three terms are just the flat Minkowski metric in spherical coordinates. Using the definition of $r$ given above to transform into `Cartesian' coordinates the metric becomes
\begin{equation}
  ds^2=\eta_{\mu\nu}dx^\mu dx^\nu+\frac{2GM}{r}k_\mu k_\nu dx^\mu dx^\nu\, ,
\end{equation}
where the null vector $k^\mu$ is defined by
\begin{equation}
  k^\mu=\left(1,\frac{x^i}{r}\right),\qquad i=1...3.
\end{equation}
The metric is now in Kerr-Schild form:
\begin{equation}\label{kerrsh}
  g_{\mu\nu}=\eta_{\mu\nu}+\frac{2GM}{r}k_\mu k_\nu \, .
\end{equation}
In terms of a metric
\begin{equation}
 g_{\mu\nu}=\eta_{\mu\nu}+\kappa\, h_{\mu\nu}\,
\end{equation}
we have
\begin{equation}
  h_{\mu\nu}=\frac{\kappa}{2}\phi k_\mu k_\nu, \qquad \phi=\frac{M}{4\pi r}\, .
\end{equation}
The single copy then yields
\begin{equation}
  A^\mu=\frac{gc_aT^a}{4\pi r}k^\mu \, ,
\end{equation}
where, motivated by the prescription for amplitudes, we have made the replacements
\begin{equation}\label{DCr}
  \frac{\kappa}{2}\rightarrow g, \quad M\rightarrow c_aT^a, \quad k_\mu k_\nu\rightarrow k_\mu \, .
\end{equation}
The first one is just choosing the right coupling constant, the second one being the correspondence between the charges of  the two theories.

\subsubsection{Taub-NUT spacetime}
The solution known as Taub-NUT was first derived by Taub \cite{Taub:1950ez} and generalised by Newman, Tamburino, and Unti \cite{Newman:1963yy} in 1963. The NUT solution is a non-asymptotically flat, axially symmetric stationary solution.
Following Ort\'in \cite{Ortin:2015hya}, the Taub-NUT metric can be written as
\begin{equation}
  ds^2=-f(r)(dt-2N\cos{\theta}d\phi)^2+f(r)^{-1}dr^2+(r^2+N^2)d\Omega^2_{(2)},\label{Taub-NUT metric}
\end{equation}
where
\begin{equation}
  f(r)=\frac{(r-r_{+})(r-r_{-})}{r^2+N^2},\qquad r_{\pm}=M\pm r_0,\qquad r_{0}^2=M^2+N^2.
\end{equation}
This can be thought of as a generalisation of the Schwarzschild solution with an additional topological  charge. This solution exhibits the following interesting properties:
\begin{itemize}
  \item The solution is not trivial in the limit $M\rightarrow 0$.
  \item Taking the Newtonian limit shows that $M$ is indeed the mass of the source. The NUT charge has no Newtonian analogue.
  \item The solution defines its own class of asymptotic behaviour labeled by $N$ and is associated with the non-vanishing at infinity of the $g_{t\phi}$ component of the metric.
  \item The solution admits Dirac-like singularities at $\theta=0,\pi$, requiring the introduction of two coordinate patches.
\end{itemize}
We saw earlier that the mass in gravity single copies to an electric or colour charge. The NUT charge $N$ single copies to a magnetic monopole on the gauge theory side as follows.

The Taub-NUT metric written in coordinates introduced by Plebanski \cite{PLEBANSKI1975196} exhibits a double Kerr-Schild form, as shown in \cite{CHONG2005124}. This means the Taub-NUT metric may be written as:
\begin{equation}
    g_{\mu\nu}=\eta_{\mu\nu}+\kappa h_{\mu\nu}
    =\eta_{\mu\nu}+\kappa(\phi k_{\mu}k_{\nu}+\psi l_{\mu}l_{\nu}) \, ,
\end{equation}
where the vectors $k^\mu$ and $l^\mu$ must satisfy the following conditions
\begin{equation}
  k^2=l^2=k\cdot l=0\, ,\qquad (k\cdot D)k_\mu=0,\qquad (l\cdot D)l_\mu=0 \, .
\end{equation}
This metric linearises the Einstein equations as with the single Kerr-Schild form. 
The generalisation of the single copy prescription for the gauge field is then
\begin{equation}
 A_{\mu}^a=c^a(\phi k_{\mu}+\psi l_{\mu}) \, .
\end{equation}
Following \cite{Luna:2015paa}, we have made the following substitutions
\begin{equation}
  \frac{M\kappa}{2}\rightarrow (c_aT^a)g_s\, ,\qquad \frac{N\kappa}{2}\rightarrow (c_aT^a)\tilde{g}_s\, .
\end{equation}
The field strength is then
\begin{equation}
  F=\frac{1}{2}F_{\mu\nu}dx^\mu dx^\nu=-\frac{c_aT^a}{8\pi}\Big(\frac{g_{s}}{r^2}dt\wedge dr+\tilde{g}_{s}\sin{\theta}d\theta\wedge d\phi\Big) \, .\label{Taub-NUT single copy}
\end{equation}
The first term on the right-hand side above is  the pure electric charge corresponding to a Coulomb solution that was derived in the Schwarzschild case. The second term is a magnetic monopole charge which is single copied from the NUT contribution on the gravity side. In summary, the Schwarzschild single copies to a Coulomb solution and the NUT charge to a magnetic monopole.

\section{Solution generating transformations in general relativity}\label{ehlerss}

A solution generating transformation is simply a recipe for obtaining a new solution to Einstein's equations from a known one. Their existence is remarkable given that Einstein's equations are a set of ten  second order coupled non-linear differential equations that are notoriously difficult to solve. Whilst explicit symmetries in Einstein's equations make some transformations straightforward, there are  ``hidden symmetries" that allow us to generate very non-trivial solutions through this transformation technique. This approach to solving Einstein's equations was studied by Buchdahl, Ehlers, Geroch and  Ernst \cite{Ernst:1967wx}. The transformations require a Killing symmetry in the spacetime. When such a symmetry is present then there are a set of solution-generating techniques which we describe below.

\subsection{Ehlers transformation}

The Ehlers transformation is a transformation that acts on the parameters of a static solution of Einstein gravitational field equations and generates other solutions of the field equations that are in general stationary. In what follows we describe how the Ehlers transformation works \cite{Ehlers:1961zza}, following \cite{Momeni:2005uc}.
It is assumed that the spacetime possesses a time-like Killing vector $\xi$ that generates an isometry. First perform a (1+3) decomposition of the metric and choose coordinates $x^{\mu}=\{x^0,x^i\}$ to put the line element in the following form
\begin{equation}
ds^2=-e^{2U}(dx^0+A_{i}dx^{i})^2+d\ell^2\, ,
\end{equation}
where  we define
\begin{equation}
A_{i}=\frac{-g_{0i}}{g_{00}} \, , \quad 
e^{2U}= -g_{00}\, , \quad 
dl^2 = \gamma_{ij}dx^i dx^j
\end{equation}
and
\begin{equation}
  \gamma_{\alpha\beta}=(-g_{\alpha\beta}+\frac{g_{0\alpha}g_{0\beta}}{g_{00}})  \, .
\end{equation}
The Ehlers transformation states that if
\begin{equation}\label{Ehler form}
  g_{\mu \nu}dx^\mu dx^\nu=-e^{2U}(dx^0)^2+e^{-2U}d\tilde{\ell}^2\, ,
\end{equation}
with $d\tilde{\ell}^2=e^{2U}d\ell^2$, is the metric of a static spacetime, then
\begin{equation}\label{result form}
  \bar{g}_{\mu \nu}dx^\mu dx^\nu=-(\alpha\cosh(2U))^{-1}(dx^0+A_{i}dx^{i})^2-\alpha\cosh(2U)d\tilde{\ell}^2\, ,
\end{equation}
where $\alpha$ is a positive constant, $U=U(x^i)$ and $A_i=A_{i}(x^j)$, is the metric of a stationary spacetime provided that $A_{i}$ satisfies Ehlers equation
\begin{equation}\label{Ehlers eq2}
 -\alpha \sqrt{\tilde{\gamma}}\,\epsilon_{ijk}\, U^{,k}=A_{[i,j]}\, ,
\end{equation}
where $\tilde{\gamma}_{ab}$ is the conformal spatial metric. Using this method, one generates a stationary solution from a static one by finding the potential $A_{a}$ which is related to a given static potential $U$. Let us see how this comes about by applying this procedure to the Schwarzschild metric.

Start with the Schwarzschild metric in the form \eqref{Ehler form}, with the potential taking the form
\begin{equation}\label{Pot}
U_c=\frac{1}{2}\ln{(1-\frac{2m}{r})}+C
\end{equation}
and
\begin{equation}
  d\tilde{\ell}^2=dr^2+r^2e^{2U}(d\theta^2+\sin^2{\theta}d\phi^2)\, ,
\end{equation}
where $C$ is a constant and $U(x^a)$ only depends on the spatial coordinates. Then the Ehlers equation \eqref{Ehlers eq2} reads
\begin{equation}
  -\alpha r^2e^{2U}\sin{\theta}\,\tilde{\gamma}^{rr}U_{,r}=\frac{1}{2}(A_{\theta,\phi}-A_{\phi,\theta}) \, .
\end{equation}
One may choose a gauge such that $A_{\theta,\phi}=0$. Then substituting \eqref{Pot} in the above equation and solving for the field $A_{\phi}$ we find
\begin{equation}\label{gravmag}
  A_{\phi}(\theta)=-2\,\alpha\, m \cos{\theta}\, .
\end{equation}
The resulting metric is then given by  \eqref{result form} as
\begin{equation}\label{momenehler}
ds^2=-\frac{r(r-2m)}{\alpha f(r)}(dt-2\alpha m\cos{\theta}d\phi)^2+\frac{\alpha f(r)}{r(r-2m)}dr^2+\alpha f(r)d\Omega^2,
\end{equation}
where
\begin{equation}
  f(r)=r^2\left(\frac{1+c_1^2}{2c_1}\right)+2m^2c_1-2mrc_1\, , \quad c_1=e^{2C} \, .
\end{equation}

Recovering the Schwarzschild metric as $\alpha\rightarrow 0$ requires us to set
\begin{equation}
  \alpha=\frac{2c_1}{1+c_1^2}, \quad |c_1|\leq 1.
\end{equation}
Now apply the following changes of variables and redefinitions of constants;
\begin{equation}\label{massc}
  M=m\,\frac{1-c_1^2}{1+c_1^2}\, , \quad \alpha\, m=N, \quad r-N c_1=R\, ,
\end{equation}
to get
\begin{equation}
  ds^2=-\frac{R^2-2MR-N^2}{R^2+N^2}(dt-2N\cos{\theta}d\phi)^2
  +\frac{R^2+N^2}{R^2-2MR-N^2}dR^2
  +(R^2+N^2)d\Omega^2 \, ,
\end{equation}
which is the metric for the NUT spacetime with NUT charge $N$. Remarkably, the constant $C$ in the potential $U$, while having no effect on the Schwarzschild spacetime, plays an important role in generating solutions when the Ehlers transformation is applied. Had we started with $C=0$ we would have ended up with the pure NUT spacetime ($M=0$). This means that the seed metric for both the NUT with mass and pure NUT metric is the Schwarzschild metric.

%%%%%%%%%%%%%%%%%%%%%%%%%%%%%%%%%%%%%%%%%%%%%

\subsection{Buchdahl's reciprocal transformation}
In \cite{PhysRev.115.1325} Buchdahl showed that if a solution of Einstein's equations admits a Killing vector field one can obtain a new solution by applying the so called ``reciprocal transformation''. For coordinates adapted to the Killing direction such that the Killing vector is given by $\frac{\partial}{\partial x^\alpha}$, then the line element may be written
\begin{equation}
  ds^2=g_{\beta\gamma}dx^{\beta}dx^{\gamma}+g_{\alpha\alpha}(dx^\alpha)^2 \, .
\end{equation}
The reciprocal transformation then generates the following line element
\begin{equation}
  ds^2=g_{\alpha\alpha}^{\frac{2}{d-3}}g_{\beta\gamma}dx^{\beta}dx^{\gamma}+g_{\alpha\alpha}^{-1}(dx^\alpha)^2 \, .
\end{equation}
Thus transformation may be written acting on metric components in this adapted coordinate system as
\begin{equation}
  (g_{\alpha\alpha},g_{\beta\gamma})\rightarrow (g_{\alpha\alpha}^{-1},g_{\alpha\alpha}^{\frac{2}{d-3}}g_{\beta\gamma})\, .
\end{equation}
The reader familiar with string theory will note that Buchdahl's reciprocal transformation is essentially T-duality  as a solution generating transformation,  predating it by some thirty years.
We shall apply this to the Schwarzschild metric. As noted above, the $d=4$ Schwarzschild metric admits a time-like Killing vector $\partial/\partial t$ so that the transformation yields
\begin{equation}
  (g_{tt},g_{ij})\rightarrow (g_{tt}^{-1},g_{tt}^2g_{ij})\, ,
\end{equation}
which upon substitution into the metric gives
\begin{equation}\label{timelikebuch}
  ds^2=-\frac{dt^2}{1-\frac{2M}{r}}+\Big(1-\frac{2M}{r}\Big) dr^2+r^2\Big(1-\frac{2M}{r}\Big)^2(d\theta^2+\sin{\theta}^2d\phi^2).
\end{equation}
The Schwarzschild metric also admits the spatial Killing vector $\partial/\partial\phi$. Applying the same transformation, but now with $\alpha=\phi$, we get the following metric
\begin{equation}
  ds^2=-r^4\sin^4{\theta}\Big(1-\frac{2M}{r}\Big)  dt^2+r^4\sin^4{\theta}\Big(1-\frac{2M}{r}\Big) ^{-1} dr^2  +r^6\sin^4{\theta}d\theta^2
  +\frac{1}{r^2\sin^2{\theta}}d\phi^2\, .
\end{equation}
This new solution is completely unrelated to the original seed metric, unlike \eqref{timelikebuch} which was obtained using the time-like Killing symmetry. However, the metric \eqref{timelikebuch} is related to the Schwarzschild metric by a simple coordinate transformation  $R=r-2M$ whereby one obtains the Schwarzschild metric but with negative mass parameter
\begin{equation}
  ds^2=-\Big(1-\frac{2M}{R}\Big) dt^2+\Big(1+\frac{2M}{R}\Big)^{-1}dR^2 +R^2(d\theta^2+\sin{\theta}^2d\phi^2).
\end{equation}

If we write the negative mass Schwarzschild metric in in Kerr-Schild form
\begin{equation}
g_{\mu\nu}=\eta_{\mu\nu}-\frac{2GM}{r}k_\mu k_\nu
\end{equation}
and then do the usual single copy procedure where $-M\rightarrow -c_aT^a$ we arrive at
\begin{equation}
  \tilde{A}_\mu=(-\frac{gc_aT^a}{4\pi r},0,0,0)\, ,
\end{equation}
which has the opposite sign of the gauge field $A_{\mu}$ compared to the single copy of the positive mass Schwarzschild solution \cite{Monteiro:2014cda}, {\it i.e.}
\begin{equation}
  \textrm{`Single copied 4D Buchdahl'}: A_{\mu} \rightarrow -{A_\mu}\, .
\end{equation}
This shows that the Buchdahl reciprocal transformation associated with the time-like Killing vector in the Schwarzschild spacetime is the gravitational analogue of charge conjugation on the gauge theory side.

We now examine this transformation acting on the Taub-NUT solution by first using the Schwarzschild metric as our seed metric on which we act with the Buchdahl transform, followed by the Ehler transformation. Firstly, note that Buchdahl transformation corresponds to sending $U\rightarrow \hat{U}=-U$ when one writes the Schwarzschild metric in the form
\begin{equation}
  ds^2=-e^{2U}(dx^0)^2+e^{-2U}d\tilde{\ell}^2, \quad U=\frac{1}{2}\ln(1-\frac{2M}{r})\, .
\end{equation}
Thus the Buchdahl-transformed Schwarzschild metric reads
\begin{equation}
    ds^2=-e^{-2U}(dx^0)^2+e^{2U}d\hat{\ell}^2, \quad d\hat{\ell}^2=dr^2+e^{-2U}r^2d\Omega^2.
\end{equation}
which as before is just the Schwarzschild metric with negative mass as shown earlier. Now construct the Ehlers transformed metric as before. Considering the Ehlers equation for the reciprocal solution
\begin{equation}
  -\alpha \sqrt{\gamma}\epsilon_{\alpha\beta\gamma}\hat{U}^{,\gamma}=\alpha \sqrt{\gamma}\epsilon_{\alpha\beta\gamma}U^{,\gamma}=\hat{A}_{[\alpha,\beta]}
\end{equation}
and comparing it to the standard one \eqref{Ehlers eq2} we find
\begin{equation}
\hat{A}_\mu=-A_\mu\, .
\end{equation}
Then, following exactly the same steps as before, {\it i.e.} solving for $A_\mu$ while imposing axial symmetry, we find
\begin{equation}
\hat{A}_\phi(\theta)=2\alpha M \cos{\theta}
\end{equation}
(up to a constant), and as before the NUT charge is given by
\begin{equation}
  \hat{N}=\alpha M=-N \, .
\end{equation}
This has acquired a negative sign, comparing the field $A_\mu$ to Eq. \eqref{gravmag}. The reciprocal Taub-NUT metric then reads
\begin{equation}\begin{split}
  &ds^2=-\big(\alpha\cosh(2U)\big)^{-1}(dx^0+\hat{A}_{\beta}dx^{\beta})^2+\alpha\cosh(2U)d\hat{\ell}^2\, , \\
  &d\hat{\ell}^2=dr^2+e^{-2U}r^2d\Omega^2.
\end{split}
\end{equation}
Upon applying the single copy procedure \cite{Luna:2015paa} as in the previous sections using the multi-Kerr-Schild form of the metric one finds again that the resulting gauge field has negative electric and magnetic monopole charges.

In summary, we show below diagrams of the transformations and their effects on the parameters of both the Schwarzschild and Taub-NUT solutions. The vertical lines indicate the application of the double/single copy (``S.D.C") procedure.
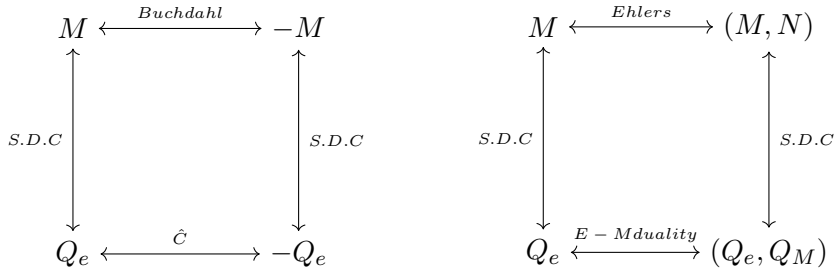
\begin{center}
\begin{tikzpicture}[scale=3]
\node (A) at (0,1) {$M$};
\node (B) at (1,1) {$-M$};
\node (C) at (0,0) {$Q_e$};
\node (D) at (1,0) {$-Q_e$};
\path[<->,font=\tiny]
(A) edge node[above]{$Buchdahl$} (B)
(A) edge node[left]{$S.D.C$} (C)
(B) edge node[right]{$S.D.C$} (D)
(C) edge node[above]{$\hat{C}$} (D);
\end{tikzpicture}
\hspace{1cm}
\begin{tikzpicture}[scale=3]
\node (A) at (0,1) {$M$};
\node (B) at (1,1) {$(M,N)$};
\node (C) at (0,0) {$Q_e$};
\node (D) at (1,0) {$(Q_e,Q_M)$};
\path[<->,font=\tiny]
(A) edge node[above]{$Ehlers$} (B)
(A) edge node[left]{$S.D.C$} (C)
(B) edge node[right]{$S.D.C$} (D)
(C) edge node[above]{$E-M duality$} (D);
\end{tikzpicture}
\end{center}

From the action of the Ehlers transformation on the charges in the double copy one sees that the single copy of the Ehlers transformation is electromagnetic duality in the gauge theory;  alternatively the double copy of electromagnetic duality is generated by the Ehlers transformation in gravity.

More formally, the Ehlers transformation is an element of $SL(2,\mathbb{R})$ as introduced by Geroch \cite{Geroch}. However when this acts on the Taub NUT solution the quantisation of NUT charge means that the group must be broken to $SL(2,\mathbb{Z})$ so as to preserve the quantisation condition. This exactly follows what happens with the electromagnetic duality group acting on the dyon spectrum. Classically the group  is $SL(2,\mathbb{R})$ but this reduces to $SL(2,\mathbb{Z})$ in order to maintain Dirac quantisation for the magnetic charges.

%%%%%%%%%%%%%%%%%%%%%%%%%%%%%%%%%%%
%%%%%%%%%%%%%%%%%%%%%%%%%%%%%%%%%%%
%%%%%%%%%%%%%%%%%%%%%%%%%%%%%%%%%%%
%%%%%%%%%%%%%%%%%%%%%%%%%%%%%%%%%%%

\section{The Tensor Weyl Double Copy}

\subsection{Definitions}
The double copy as described above is defined in terms of the gauge connection and the metric. It is a natural question to ask whether one might study a double copy directly in terms of field strengths and curvatures. This was investigated in \cite{Luna:2018dpt}, where a particularly nice form of the double copy was obtained using spinors -- writing the spinor corresponding to the Weyl tensor in terms of the spinor for the Maxwell tensor as
\begin{equation}
\label{spinorDC}
C_{ABCD} = \frac{1}{S} f_{(AB}f_{CD)} \, ,
\end{equation}
where $S$ is a suitable function related to the ``zeroth copy". This ``Weyl double copy" was shown to be consistent with the previously known Kerr-Schild double copy, and resolved some of ambiguities in that formulation. It also presented new double copy interpretations of the Eguchi-Hanson instanton, and the C-metric, relating the latter to the Li\'enard-Wiechert potential for a pair of uniformly accelerated charges. Extending this four-dimensional result to higher dimensions requires an appropriate study of spinors and curvature invariants in higher dimensions, and the latter has been explored recently in \cite{Monteiro:2018xev}.
\\

A higher-dimensional Weyl double copy might also be investigated in terms of a tensor version of the spinor Weyl double copy. One can obtain this by translation from the spinor equations of course;  more directly one can note that this must involve writing the Weyl tensor as a sum of terms quadratic in the Maxwell tensor. Keeping in mind the algebraic symmetries of the Weyl tensor, in four dimensions there are two independent expressions  that one can write down:
\begin{align}
\label{CDdefns}
C_{\mu\nu\rho\sigma} [F] &= F_{\mu\nu}F_{\rho\sigma} - F_{\rho\mu}F_{\nu\sigma}-3g_{\mu\rho}{F_\nu}^\lambda F_{\sigma\lambda} + \frac{1}{2}g_{\mu\rho}g_{\nu\sigma}F^2   \; \Big\vert_s  \, ,
\\
D_{\mu\nu\rho\sigma} [F] &= \frac{1}{2}\left( F_{\mu\nu}\Ft_{\rho\sigma} - F_{\rho\mu}\Ft_{\nu\sigma}-3g_{\mu\rho}{F_\nu}^\lambda \Ft_{\sigma\lambda} + \frac{1}{2} g_{\mu\rho}g_{\nu\sigma}F.\Ft \right) + (F \leftrightarrow \Ft)   \;  \Big\vert_s  \, ,
\end{align}
where $F^2= F^{\lambda\delta}F_{\lambda\delta}$,   $F.\Ft =F^{\lambda\delta}\Ft_{\lambda\delta}$, and $\Ft_{\mu\nu} =\frac{1}{2}\sqrt{g}\, \epsilon_{\mu\nu\rho\sigma}F^{\rho\sigma}$ with  $\epsilon_{\mu\nu\rho\sigma}$ the numerical alternating symbol.
In equations like those above, the symbol ``$\vert_s$" applies to the expression on the right-hand side of the equation, and it means to anti-symmetrise in the indices $\mu,\nu$ and in $\rho, \sigma$, with unit weight.
In $D$  dimensions there is no equivalent of  $ D_{\mu\nu\rho\sigma} [F]$ and one just has
\begin{equation}
\label{Dhigher}
C^{(D)}_{\mu\nu\rho\sigma} [F] = F_{\mu\nu}F_{\rho\sigma} - F_{\rho\mu}F_{\nu\sigma}-\frac{6}{D-2}g_{\mu\rho}{F_\nu}^\lambda F_{\sigma\lambda} + \frac{3}{(D-1)(D-2)}g_{\mu\rho}g_{\nu\sigma}F^2   \; \Big\vert_s  \, .
\end{equation}
\\
We make some comments on the higher-dimensional double copy at the end of the paper and hereon consider four dimensions. A four-dimensional Weyl tensor double copy must  involve a linear sum of the two expressions $C_{\mu\nu\rho\sigma} $ and $D_{\mu\nu\rho\sigma} $, with suitable coefficients which will in general be functions of the relevant variables and constants.
We note some useful properties of these expressions:
 $C_{\mu\nu\rho\sigma} [\Ft] = C_{\mu\nu\rho\sigma} [F]$, $\tilde C_{\mu\nu\rho\sigma} [F] = D_{\mu\nu\rho\sigma} [F]$, $D_{\mu\nu\rho\sigma} [\Ft] = -D_{\mu\nu\rho\sigma} [F] $ and
\begin{equation}
C_{\mu\nu\rho\sigma} [aF+b\Ft] = (a^2+b^2)\,C_{\mu\nu\rho\sigma} [F] + 2ab\, D_{\mu\nu\rho\sigma} [F]
\end{equation}
for any coefficients $a,b$. Define the self-dual and anti-self-dual parts of a two form $F$ via $F^\pm ={1\over 2}(F \pm \Ft)$ with a similar formula for $C^\pm$, with the action either on the left or right pair of indices. Then
\begin{equation}
C_{\mu\nu\rho\sigma} ^\pm[a F+b\Ft] = C_{\mu\nu\rho\sigma} [(a\pm b)F^\pm] \, .
\end{equation}

%%%%%%%%%%%%%%%%%%%%%%%%%%%%%%%%%

\subsection{Ehlers and the double copy: the Schwarzschild case}

We would like to explore how the Ehlers transformation described earlier may be understood in terms of the double copy. Consider as starting point the Schwarzschild metric. We will find it useful to work in coordinates $(u,v,p,q)$ with the metric
\begin{align}
\label{Schw_metric}
ds^2 &=  \frac{1}{(1-pq)^2} \Bigg[2 i (du+q^2 dv) dp -2(du-p^2 dv) dq   + \frac{2mp^3}{(p^2+q^2)} (du+q^2 dv)^2   \nonumber \\
&\qquad\qquad\qquad+ \frac{2mq}{(p^2+q^2)}(du-p^2 dv)^2 \Bigg]   \, .
\end{align}
The single copy Maxwell tensor is then \cite{Luna:2018dpt}
\begin{equation}
F_S = {e\over (p^2+q^2)^2} \Big[ 2 pq (du+q^2dv)dp - (p^2-q^2) (du-p^2dv)dq \Big]  \, .
\end{equation}
It is then straightforward to show that the Weyl tensor $C_S$ for the Schwarzschild metric is given by
\begin{equation}
C^S_{\mu\nu\rho\sigma}  = -{4\over (1-pq) e} \Big( q \,C_{\mu\nu\rho\sigma} [F_S] -i p\,D_{\mu\nu\rho\sigma} [F_S] \Big) \,\Big\vert_{e\rightarrow m}  \, ,
\end{equation}
where $\vert_{e\rightarrow m}$ means to replace $e$ by $m$ on the right-hand side of the equation. This result may be written simply as
\begin{equation}
\label{SchwWeyl}
C^S_{\mu\nu\rho\sigma} = -{4\over (1-pq)} \Big( C_{\mu\nu\rho\sigma} [ \alpha_S F_S^+ + (c.c)]\Big) \,\Big\vert_{e\rightarrow m}  \, ,
\end{equation}
where $\alpha_S = \sqrt{{-i(p+iq)\over e}}$.

Now let us consider the Taub-NUT metric in the corresponding coordinate system:
\begin{align}
\label{TNmetric}
ds^2 &=  \frac{1}{(1-pq)^2} \Bigg[2 i (du+q^2 dv) dp -2(du-p^2 dv) dq   +2p \frac{mp^2+n}{(p^2+q^2)} (du+q^2 dv)^2   \nonumber \\
&\qquad\qquad\qquad+ 2q\frac{m+nq^2}{(p^2+q^2)}(du-p^2 dv)^2 \Bigg]   \, ,
\end{align}
where $n$ is the Taub-NUT charge. The single copy Maxwell tensor in this case is
\begin{equation}
\label{MaxwellTN}
F_T = {1\over(p^2+q^2)^2} \Big[ \big(2 e p q +g(p^2-q^2) \big)(du +q^2 dv) dp
         +  \big(2 g p q -e(p^2-q^2) \big) (du-p^2dv) dq    )   \Big]  \, .
\end{equation}
This can be  expressed simply in terms of the Schwarzschild Maxwell single copy tensor as
\begin{equation}
F_T = F_S - {i g\over e}\Ft_S  \, .
\end{equation}
Now, if we  make the replacements
\begin{equation}
\label{Ehlers}
F_S \rightarrow  F_S - \frac{ ig}{e} \Ft_S, \qquad e\rightarrow e-i g\, ,
\end{equation}
on the right-hand side of
\eqref{SchwWeyl}, and then make the replacements
$e\rightarrow m, g\rightarrow n$ then we find a tensor that we will call $C^T$ which is
\begin{equation}
\label{TNWeyl}
C_{\mu\nu\rho\sigma} ^T := -{4\over (1-pq)} \Big( C_{\mu\nu\rho\sigma} [ \alpha _T F_T^+ + (c.c)]\Big) \Big\vert_{e\rightarrow m, g\rightarrow n} \, ,
\end{equation}
where $\alpha_T = \sqrt{{-i(p+iq)\over (e-ig)}}$.
It can be checked that $C^T$ is the Weyl tensor for the Taub-NUT metric   \eqref{TNmetric}. (Note that we implicitly assumed that the charge $e$ is complex prior to the shift, and that in going from \eqref{SchwWeyl} to \eqref{TNWeyl} the metric dependence in $C_{\mu\nu\rho\sigma} [F]$ also needs to shift from \eqref{Schw_metric} to  \eqref{TNmetric}.)
\\

Thus we see that the Ehlers transformation which takes one from the Schwarzschild to the Taub-NUT spacetime can be seen via the Weyl double copy as a simple duality transformation \eqref{Ehlers} (combined with identifying $(e,g)$ with $(m,n)$) which maps between the two Weyl double copy curvatures. It is instructive to return to the spinor form of the Weyl double copy \eqref{spinorDC} in the light of this (see Section 4 of \cite{Luna:2018dpt}). The transformation $eF\rightarrow (e-ig)F$ induces the shifts $eF^\pm\rightarrow (e\mp ig)F^\pm$. The Maxwell field strength spinor $f_{AB}$ depends only on the self-dual part of the Maxwell tensor and thus transforms according to this formula. The scale function $e\, S$ in \eqref{spinorDC}  transforms to $(e-ig)S$ and hence the double copy formula yields
\begin{equation}
\label{WeylSpinorTransform}
m\, C_ {ABCD}\rightarrow  (m-in)\, C_ {ABCD}\, ,
\end{equation}
correctly mapping the Schwarzschild Weyl spinor to the Taub-NUT one.

%%%%%%%%%%%%%%%%%%%%%%%%%%%%%%%%%

\subsection{Type D metrics}\label{subsec: Type D metrics}

The Taub-NUT example considered above is a special case of the general vacuum type D solution with vanishing cosmological constant \cite{Plebanski:1976gy}, as given in  \cite{Luna:2018dpt}:
\begin{equation}
\label{typeDmetric}
ds^2 = {1\over(1-pq)^2}  \Bigg[2 i (du+q^2dv) dp - 2(du-p^2dv)dq +  \frac{P(p)}{p^2+q^2}\,(du+q^2dv)^2
-  \frac{Q(q)}{p^2+q^2}\,(du-p^2dv)^2  \Bigg] \, ,
\end{equation}
with
\begin{align}
P(p) =  \gamma (1-p^4) +2 n p- \epsilon p^2 +2 m p^3 \,, \nonumber \\
Q(q) =  \gamma (1-q^4) - 2m q + \epsilon q^2 -2 n q^3 \,,
\end{align}
where the parameters $m,n,\gamma,\epsilon$ are related to the mass, NUT charge, angular momentum and acceleration (see \cite{Griffiths:2005qp} for a discussion of the various limits and definitions which enable the identifications in different cases).

The single copy Maxwell tensor in this case is the same as the one for the Taub-NUT metric \eqref{TNmetric}. It is then natural to investigate the Weyl double copy in this case and, indeed, one finds that the Weyl tensor $C^D$ for the type D metric \eqref{typeDmetric} is  given by the same formula as that for the TN case:

\begin{equation}
\label{TypeDWeyl}
C^D_{\mu\nu\rho\sigma} := -{4\over (1-pq)} \Big(C_{\mu\nu\rho\sigma} [ \alpha _T F_T^+ + (c.c)]\Big) \, \Big\vert_{e\rightarrow m, g\rightarrow n}  \, ,
\end{equation}
with $\alpha_T = \sqrt{{-i(p+iq)\over (e-ig)}}$. Note that the metric \eqref{typeDmetric}  enters the right-hand side of  \eqref{TypeDWeyl} so that this doesn't simply reproduce $C_{\mu\nu\rho\sigma} ^T$.

One can then ask if an Ehlers transformation will take one from the spacetime with Type D metric \eqref{typeDmetric} with $n = 0$, to that with nonzero $n$. To see this, consider the type D metric $g_{D_0}$ with vanishing NUT charge. This satisfies
\begin{equation}
\label{eq::TypeDWeylNoNUT}
C^{D_0}_{\mu\nu\rho\sigma} = -{4\over (1-pq)} \Big(C_{\mu\nu\rho\sigma} [ \alpha _S F_S^+ + (c.c)]\Big) \, \Big\vert_{e\rightarrow m}  \, .
\end{equation}
Then if we make the replacements $F_S \rightarrow  F_ S- {i g\over e}\Ft_ S$ and $e\rightarrow e-i g$ in the right-hand side of the above, and shift the metric from $g_{D_0}$ to \eqref{typeDmetric}, we find that we reproduce  \eqref{TypeDWeyl}.

%%%%%%%%%%%%%%%%%%%%%%%%%%%%%%%%%%%%%%%%%%%%%%%%%%%%%%%

\section{The $SL(2,\mathbb{R})$ transformations}

\subsection{The spacetime Ehlers group}

We would now like to  discuss how   $SL(2,\mathbb{R})$ transformations  act more generally in the context of the double copy.
For this purpose, it will be useful to use a generalisation of the Ehlers procedure for  more general Killing vector fields: the \textit{spacetime Ehlers group} \cite{Mars:2001gd}, which we now summarise briefly.\footnote{We comment that this paper and  work following from it (see for example \cite{Mars:2016ynw} and references therein) anticipate some of the formul\ae\ of the double copy -  e.g., the vanishing of the Mars-Simons tensor defined below corresponds to the self-dual part of the tensor double copy, and the spinor form of this can  be found in \cite{Gomez-Lobo:2016ykv}.}
Given a Killing vector field $\xi=\xi^\mu\partial_\mu$ and one-form $W=W_\mu dx^\mu$ on a Lorentzian manifold with metric $g_{\mu\nu}$, satisfying the vacuum Einstein equations, the spacetime Ehlers group is defined in \cite{Mars:2001gd} by the transformation
\begin{equation}
\label{newmetric}
g_{\mu\nu}\rightarrow \Omega^2 g_{\mu\nu}-2\xi_{(\mu}W_{\nu)}-\frac{\lambda}{\Omega^2}W_\mu W_\nu,
\end{equation}
where $\lambda=-\xi^\mu\xi_\mu$ and $\Omega^2\equiv\xi^\mu W_\mu+1\geq 1$ with the inequality holding over the whole geometry.
Define  the twist potential $\omega_\mu=\sqrt{-\operatorname{det}(g)}\,\epsilon_{\mu\nu\sigma\rho}\xi^\nu\nabla^\sigma\xi^\rho$
and the Killing tensor two form $F_{\mu\nu}=2\partial_{[\mu}\xi_{\nu]}$ (note that we have a factor of $2$ here, and a factor of $1/2$ in the definitions of the (anti-)self-dual parts of $F$, in comparison with \cite{Mars:2001gd}).
Then the Ernst one-form
\begin{equation}
\sigma_\mu :=  2 \xi^\nu F^+_{\nu\mu} =\nabla_\mu\lambda-i\omega_\mu \,
\end{equation}
is closed, following from the vanishing of the Ricci tensor, and so locally
$\sigma_\mu =   \nabla_\mu\sigma \label{ernstoneform2}$
for some complex function $\sigma$.\footnote{This satisfies $\nabla^2\sigma=-(F^+_{\mu\nu})^2$. In this construction we do not see a direct emergence of the standard harmonic \lq\lq zeroth copy" function, as in the Kerr-Schild formulation. }
\\
The spacetime Ehlers group is then defined for $W$ satisfying
\begin{subequations}\label{Wform}
\begin{gather}
2\nabla_{[\mu}W_{\nu]}=-4\gamma\,{\rm Re}[(\gamma \bar\sigma+i\delta)\,F^+_{\mu\nu}]\label{Wforma}\, ,\\
\xi^\mu W_\mu+1 = (i\gamma\sigma+\delta)(-i\gamma \bar{\sigma}+\delta) \label{Wformb}\, ,
\end{gather}\end{subequations}
where a bar denotes complex conjugation, and $\gamma$ and $\delta$ are non-simultaneously vanishing real constants, which as a pair fix the gauge of $W$. After repeated action, the transformation defines an $SL(2,\mathbb{R})$ group action on the Ernst scalar by the M\"obius map
\begin{equation}
\label{sl2r}
\sigma\rightarrow\frac{\alpha\sigma+i\beta}{i\gamma\sigma+\delta},\qquad\text{where}\quad \beta\gamma+\alpha\delta=1.
\end{equation}
The self-dual part of the Killing tensor transforms as
\begin{equation}
\label{Ftransform}
F^+_{\mu\nu} \rightarrow \frac{1}{(i\gamma\sigma+\delta)^2}\Big(\Omega^2 F^+_{\mu\nu} -W_{[\mu} \sigma_{\nu]}  \Big)\, ,
\end{equation}
where $W, \sigma$ are the one-forms defined above. The self-dual part of the Weyl tensor transforms as
\begin{equation}
\label{Weyltransform}
C^+_{\mu\nu\rho\sigma} \rightarrow     \frac{1}{(i\gamma\sigma+\delta)^2} P^{\alpha\beta}_{\mu\nu}
P^{\gamma\delta}_{\rho\sigma}  \left( C^+_{\alpha\beta\gamma\delta}   -   \frac{6i\gamma}{i\gamma\sigma+\delta} \Big( F^+_{\alpha\beta}F^+_{\gamma\delta} -\frac{1}{3}I_{\alpha\beta\gamma\delta} (F^+)^2\Big) \right) \,  ,
\end{equation}
where in our conventions
 $I_{\mu\nu\rho\sigma}= \frac{1}{4}(g_{\mu\rho}g_{\nu\sigma} - g_{\nu\rho}g_{\mu\sigma} + \epsilon_{\mu\nu\rho\sigma})$
is the canonical metric in the space of self-dual two-forms and $P^{\alpha\beta}_{\mu\nu}   =\Omega^2\delta^\alpha_\mu \delta^\beta_\nu -
\delta^\alpha_\mu \xi^\beta W_\nu - \xi^\alpha W_\mu \delta^\beta_\nu$.
Notice that
\begin{equation}
 F^+_{\mu\nu}F^+_{\rho\sigma} -\frac{1}{3}I_{\mu\nu\rho\sigma} (F^+)^2 = \frac{2}{3}  C_{\mu\nu\rho\sigma}  [F^+]  \,
\end{equation}
in terms of the definition in \eqref{CDdefns}.
The Mars-Simons tensor is then defined as
\begin{equation}
   S_{\mu\nu\rho\sigma} =  C^+_{\mu\nu\rho\sigma}     - \frac{2}{3}Q\, C_{\mu\nu\rho\sigma}  [F^+] \, ,
\end{equation}
for a suitable function $Q$.
 Finally,  as discussed in Section 6 of  \cite{Mars:2001gd},  we note that the vanishing of the Mars-Simons tensor is maintained under the Ehlers transformation, with $Q$ transforming appropriately.
%

%%%%%%%%%%%%%%%%%%%%%%%%%%%%%%%%%%%%%%%%%%%%

\subsection{The Taub-NUT case}
Let us now consider applying these arguments in the context of the Weyl double copy  described earlier.
Starting from the Taub-NUT metric, in the real form \eqref{Taub-NUT metric}, consider the Killing vector $\xi=\partial_t$. Its associated two-form $F_{\mu\nu}=2\,\partial_{[\mu} \xi_{\nu]}$ is
\begin{align}
\label{F TN}
F&=\frac{2 M \left(r^2-N^2\right)+4 N^2 r}{\left(N^2+r^2\right)^2}\,dt\wedge dr
+\frac{4 N \cos (\theta ) \left(M \left(r^2-N^2\right)+2 N^2 r\right)}{\left(N^2+r^2\right)^2}\,dr\wedge d\phi \\ \nonumber
&\qquad +\frac{2\,N \sin (\theta ) \left(r (2 M-r)+N^2\right)}{N^2+r^2}\,d\theta\wedge d\phi
~.
\end{align}
This solves the Maxwell equations on the Taub-NUT background. The single copy of Taub-NUT was found in \cite{Luna:2015paa,Luna:2018dpt} and solves the flat-background Maxwell equations. The Ernst one-form is obtained from its definition
\begin{equation}
\sigma_\mu:=2\xi^\nu F^+_{\nu\mu}=\frac{2 (M-i\, n)}{( r-i\,n)^2}\,\delta^r_\mu~.
\end{equation}
In \cite{Mars:2001gd} it was proved that the Ernst one-form is exact, $\sigma_\mu=\partial_\mu \sigma$, and the integration constant can be chosen such that ${\rm Re}(\sigma)=-\xi^\mu\xi_\mu$, giving
\begin{equation}
\sigma=1-\frac{2 (N+i M)}{N+i\, r}~.
\end{equation}
Additionally, the fact that \eqref{Taub-NUT metric} has a Weyl double copy structure implies that
\begin{equation}
\begin{split}
C^+_{\alpha\beta\gamma\delta}&=-\frac{6}{c-\sigma}\left(F^+_{\alpha\beta}F^+_{\gamma\delta}-\frac{(F^{+})^{2}}{3}I_{\alpha\beta\gamma\delta}\right)~,\\
(F^{+})^{2}&= A(c-\sigma)^4~,\label{C+ as FF+}
\end{split}
\end{equation}
with $c=1$ and $A=-(4(M-iN))^{-1}$. Next, $W$ is found  by solving  \eqref{Wform}.
After this,  we can transform the original metric into \eqref{newmetric}
\begin{equation}
g^{\prime }_{\mu\nu}=\Omega^2\,g_{\mu\nu}-2\xi_{(\mu}W_{\nu)}+\frac{\xi^\sigma\xi_{\sigma}}{\Omega^2}W_{\mu}W_\nu~.
\end{equation}
In order to interpret this new metric, it is convenient to define polar coordinates in the parameter space
\begin{equation}
\rho=\sqrt{\delta^2+\gamma^2}~,\qquad \tan\zeta=\frac{\delta}{\gamma}~.
\end{equation}
Performing a charge redefinition and a change of coordinates 
\begin{equation}
\begin{gathered}
\begin{pmatrix}
M^\prime\\
N^\prime
\end{pmatrix}=\begin{pmatrix}
\cos 2\zeta& -\sin2\zeta \\
\sin2\zeta & \cos 2\zeta
\end{pmatrix}\begin{pmatrix}
\rho \,M\\
\rho\, N
\end{pmatrix}~,\label{New Charges}\\
t^\prime=\frac{ t}{\rho}~,\qquad r^\prime=\rho\,r+M^\prime(1-\cos 2\zeta)-N^\prime \,\sin 2\zeta~,
\end{gathered}
\end{equation}
the metric simplifies to
\begin{equation}
\begin{gathered}
ds^{\prime 2}=-f(r^\prime)(dt^\prime-2N^\prime\,\cos\theta\,d\phi)^2+\frac{dr^{\prime2}}{f(r^\prime)}+(r^{\prime 2}+N^{\prime 2})d\Omega^2_2~,\\
 f(r^\prime)=\frac{r^{\prime 2}-2M^\prime\,r^\prime-N^{\prime 2}}{r^{\prime 2}+N^{\prime 2}}~.\label{TN  transformed}
\end{gathered}
\end{equation}
Hence, it is still a member of the Taub-NUT family. The self-dual part of $F_{\mu\nu}$ transforms as \eqref{Ftransform}.
The integrated Ernst one-form transforms as
\begin{equation}
\label{sl2rTN}
\sigma^\prime=\frac{1}{\delta^2+\gamma^2}\,\frac{\delta \sigma+i\,\gamma}{i\gamma\sigma+\delta}~.
\end{equation}
After the transformation, \eqref{C+ as FF+} also holds with
\begin{equation}
c^\prime=\frac{1}{\gamma^2+\delta^2}~\qquad A^\prime=-\frac{(\delta+i\,\gamma)^4}{4(M-i\,N)}~,
\end{equation}
in agreement with (54) in \cite{Mars:2001gd}.
\\

Let us now study the implications for the single copy. The single copy of \eqref{Taub-NUT metric} can be written in flat spherical coordinates ($\tilde{t},\tilde{r},\theta,\phi$) \eqref{Taub-NUT single copy}
\begin{equation}
F_{T}=-\frac{M}{\tilde{r}^2}\,d\tilde{t}\wedge d\tilde{r}-N\,\sin\theta\,d\theta\wedge d\phi~.
\end{equation}
Hence, the single copy of the transformed space-time, on the same background reads
\begin{equation}
F_{T}^\prime=-\frac{M^\prime}{\tilde{r}^2}\,d\tilde{t}\wedge d\tilde{r}-N^\prime\,\sin\theta\,d\theta\wedge d\phi~.
\end{equation}
Using \eqref{New Charges}, it can be checked that the transformation in terms of the Ehlers group parameters is
\begin{equation}
F_{T}^\prime=\rho \cos(2\zeta)\, F_{T}+\rho \sin(2\zeta)\tilde{F}_{T}~.
\end{equation}
This corresponds to an electromagnetic duality rotation and a rescaling by $\rho$. Both transformations are contained in the electromagnetic duality, where the rescaling can be interpreted as the transformation of the gauge coupling \cite{Ortin:2015hya}. The zeroth copy is affected similarly, transforming using $M\rightarrow M^\prime$, $N\rightarrow N^\prime$, leaving the double copy structure intact.
The Weyl double copy
\begin{equation}
\label{WeylDCTN}
C^+_{\mu\nu\rho\sigma} = \frac{2}{\sigma^+}C_{\mu\nu\rho\sigma}[F^+]\, ,
\end{equation}
where $F^{'+}$ is the self-dual part of the transformed single-copy Killing tensor and  $\sigma^+ =\sigma-c= -2(N+iM)/(N+ir)$, is preserved: \eqref{WeylDCTN} transforms directly to the double copy in the transformed spacetime
\begin{equation}
\label{WeylDCTNnew}
C^{'+}_{\mu\nu\rho\sigma} = \frac{2}{\sigma^{+'}} C_{\mu\nu\rho\sigma}[F^{'+}]\, ,
\end{equation}
where $F^+$ is the self-dual part of the single-copy Killing tensor, now defined using the shifted metric (and the same Killing vector, although note of course that the co-vector differs in the new spacetime), and the transformed Ernst scalar is
\begin{equation}
\label{ErnstTNnew}
\sigma^{'+} = -\frac{2 (M-i N)}{(\gamma -i \delta ) (2 \gamma  M+N(\delta -i \gamma)  +r(i\delta-\gamma) 
  )} .
\end{equation}
\\
We see that in terms of the action on the fields, a restricted set of the $SL(2,\mathbb{R})$ transformations act in this case, and the orbit is within the Taub-NUT class of metrics. The two degrees of freedom are realised by the rotation parameter $\zeta$ and scaling $\rho$.

%%%%%%%%%%%%%%%%%%%%%%%%%%%%%%%%%%%%%%%%%%%%

\subsection{More general Type D metrics}

Now we may consider the general class of metrics given in \eqref{typeDmetric}. One has the  Killing vectors $U=(1,0,0,0)$ and $V=(0,1,0,0)$.  Defining the Killing forms  $F^U_{\mu\nu}=2\partial_{[\mu}U_{\nu]}, F^V_{\mu\nu}=2\partial_{[\mu}V_{\nu]}$, we find that  these are simply related to the
single copy Maxwell tensor \eqref{MaxwellTN} by
\begin{equation}
\label{SCMaxwellUV}
F_T= \frac{1}{2} \left( F^U + i \tilde F^V\right)  \,  ,
\end{equation}
which is equivalent to  $F_T^\pm =     \frac{1}{2} F^{U\pm iV}$.
The double copy is then a formula of the form of the vanishing of a Mars-Simons tensor:
\begin{equation}
\label{MarsSimons2}
 C^+_{\mu\nu\rho\sigma} =    \frac{1}{\sigma^{U+iV}}\,C_{\mu\nu\rho\sigma} [F^{(U+iV)+}]\vert_{e\rightarrow m, g\rightarrow n}     \, ,
 \end{equation}
with the Ernst scalar given by
\begin{equation}
\label{UVErnst}
\sigma^{U+iV} =    -\frac{i(1-pq)(m-in)}{p+iq}   \,  .
\end{equation}
The metric \eqref{typeDmetric} is complex, as is the relevant Killing vector $U+iV$. The derivation of the equations \eqref{Wform} relies on a real Killing vector and so cannot be used directly in this case. Thus consider instead the real form of the metric
\begin{equation}
\label{typeDmetricreal}
ds^2 = {1\over(1-y r)^2}  \Bigg[- \frac{\Delta_r}{\Sigma}(dt+y^2d\psi)^2+ \frac{\Delta_y}{\Sigma}(dt-r^2d\psi)^2 +  \frac{\Sigma}{\Delta_r} dr^2  +  \frac{\Sigma}{\Delta_y} dy^2  \Bigg] \, ,
\end{equation}
with $\Sigma=r^2+y^2$ and
\begin{align}
\Delta_r = k (1-r^4) - 2m r + \epsilon r^2 -2 n r^3 \,, \nonumber \\
\Delta_y = k (1-y^4) +2 n y- \epsilon y^2 +2 m y^3 \, .
\end{align}
Defining the single-copy Maxwell tensor
\begin{equation}
\label{typeDMaxwellreal}
F = \frac{1}{(r^2+y^2)^2} \Big[ \big(2nry-m(y^2-r^2)\big)\big(dt+y^2d\psi\big) dr +\big(2mry+n(y^2-r^2)\big)\big(dt-r^2d\psi\big) dy \Big]
 \, ,
\end{equation}
then the Weyl double copy is given by
\begin{equation}
\label{typeDWeylDCreal}
C_{\mu\nu\rho\sigma}^+ = \frac{16(iy-r)}{(m-in)(1-ry)} C_{\mu\nu\rho\sigma}[F^+]
  \,
\end{equation}
and its conjugate.
\\

Now consider the Killing vector $K = \partial_t$. The solution of the equations \eqref{Wform} here is more involved: these are solved by $W= (W_t,0,0,W_\psi) - (1-\delta^2,0,0,0)$ with
\begin{align}
W_t = &\frac{\gamma}{(r y-1)^4 (r^2+y^2)}     \Big[ 4 \gamma  m^2 (y^4+1)+4 \gamma
    n^2 (r^4+1)    \nonumber \\ &
    +4 n \Big(-\gamma  e (r^3+y)+\gamma  k (r^5+r^3 y^2-3 r^2
   y+2 r-y^3)+\delta  r (r y-1)^3\Big)    \nonumber \\ &
   + \Big((r^2+y^2) (\gamma  e^2+2 \gamma  e k
   (y^2-r^2)+ \gamma  k^2 (r^4+2 r^2 y^2-8 r y+y^4+4)+4 k
   \delta  (r y-1)^3\Big)    \nonumber \\ &
   -  4 m \Big(\gamma  e (r+y^3)-\gamma  k
   r^3+\gamma  k r^2 y^3-3 \gamma  k r y^2+\gamma  k y^5+2 \gamma  k y-2 \gamma  n
   (r^2+y^2) \nonumber \\ &
   +\delta  r^3 y^4-3 \delta  r^2 y^3+3 \delta  r y^2-\delta
   y\Big) \,
\end{align}
and
\begin{align}
W_\psi = &
\frac{2 \gamma }{(r y-1)^3 (r^2+y^2)}
 \Big(
  2 \gamma  k^2 (r^4-y^4)-2 \gamma  m^2 r y^3-2 \gamma  m^2 y^2 \nonumber \\ &
  +  k \big(-2 \gamma  e r y (r^2+y^2)+2 \gamma  m y^2
   (r^3+r y^2+2 y)+\delta  (r^5 y-r^4-r y^5+y^4)\big)  \nonumber \\ &
   +2 n r \big(\gamma  (-e) r y+\gamma  k r (r^2 y+2 r+y^3)+\delta  y (r^4
   y-r^3+2 r^2 y^3-3 r y^2+y)\big)    \nonumber \\
   & +2 \gamma  e m r y^2-2 \delta  e r^4 y^2+3
   \delta  e r^3 y-\delta  e r^2-2 \delta  e r^2 y^4+3 \delta  e r y^3-\delta  e y^2  \nonumber \\ &
   +4 \delta  m
   r^4 y^3-6 \delta  m r^3 y^2+2 \delta  m r^2 y^5+2 \delta  m r^2 y-2 \delta  m r y^4+2
   \gamma  n^2 r^2 (r y+1) \Big)\, ,
\end{align}
with the Killing tensor here  $F_{\mu\nu} = 2\partial_{[\mu}K_{\nu]}$ and
\begin{equation}
\sigma = \frac{1}{(r-iy)(1-ry)^2} \Big(
(r-i y) \big(e+k (-r^2-2 i r y+y^2+2 i)\big)+2 i m
   (y^2+i)-2 n (r^2-i)
      \Big) \, .
\end{equation}
The shifted Maxwell tensor is given by \eqref{Ftransform} and the new metric by \eqref{newmetric} with $W$ and $\sigma$ given by the expressions above.
Whilst this will lead to an $SL(2,\mathbb{R})$ action on the original solution, we note that the single-copy Maxwell tensor is not equal to the Killing tensor obtained from the Killing vector $K=\partial_t$. It would appear from this that one can either have the double copy arising from a complex Killing vector but without the $SL(2,\mathbb{R})$ action as described in \cite{Mars:2001gd}, or can keep the group action based on a real Killing vector but without the double copy being preserved.  We have seen above that this issue does not arise for the Taub-NUT solution, which is a certain limit of the metric \eqref{typeDmetric}; in addition, some more recent work has  studied complex Killing vectors and their Mars-Simon tensors \cite{Paetz:2017lkn, Paetz:2017cai}. It may thus be worth exploring this case further.

%%%%%%%%%%%%%%%%%%%%%%%%%%%%%%%%%%%%%%%%%%%%

\subsection{The Eguchi-Hanson metric}

It is of interest to consider a Riemannian metric example and we turn to the Eguchi-Hanson metric
\begin{equation}
\label{EHdoubleKS}
ds^2= 2dudv-2dXdY+\frac{\lambda}{(uv-XY)^3}(vdu-XdY)^2\, ,
\end{equation}
with coordinates $(u,v,X,Y)$ and constant $\lambda$. The single-copy (self-dual) Maxwell tensor is \cite{Berman:2018hwd,Luna:2018dpt}
\begin{equation}
\label{EHMax}
F =  \frac{2 \lambda}{(uv-XY)^3} \Big( (u v+XY)(du\wedge dv - dX\wedge dY) - 2vY du\wedge dX + 2uX dv\wedge dY\Big) \, .
\end{equation}
Consider the Killing vector
\begin{equation}
\label{EHKV}
K^\mu = (u,-v,-X,Y)\,
\end{equation}
and Killing two-form
\begin{equation}
\label{Killingtwoform}
K_{\mu\nu} = 2\partial_{[\mu} K_{\nu]}\, .
\end{equation}
The single-copy Maxwell tensor is then given by
\begin{equation}
\label{EHsc}
F_{\mu\nu}= K_{\mu\nu}^+\, .
\end{equation}
We have the relations
\begin{equation}
\label{EHErnstvecs}
\sigma_\mu^+:= 2K^\nu K_{\nu\mu}^+    = \partial_\mu \sigma^+\, , \qquad  \sigma_\mu^-:= K^\nu K_{\nu\mu}^-   = \partial_\mu \sigma^- \, ,
\end{equation}
with
\begin{equation}
\label{EHErnsts}
\sigma^+ = -\frac{2\lambda}{(uv-XY)} \, ,\qquad \sigma^- = 4(uv-XY) \, .
\end{equation}

We now consider the equations \eqref{Wform} with $\sigma\rightarrow \sigma^+, \bar\sigma\rightarrow\sigma^-$ and $\xi$ the Killing vector \eqref{EHKV}.
These are solved by
\begin{align}
\label{Wshifted}
W_\mu =  -\Big( \frac{8 \lambda \gamma^2}{u v - X Y} + 2 i \gamma\delta\Big) & (v,-u,Y,-X)  -\frac{2 i \lambda\gamma\delta}{(u v - X Y)^2} (v,0,0,-X)
\nonumber\\
&-(1-8 \lambda\gamma^2-\delta^2) \left(\frac{1}{u},0,0,0\right) \, .
\end{align}
The new metric is given by \eqref{newmetric} with $W$ given by the expression above. This is a complicated expression which we will not reproduce here.
We have checked that this is Ricci-flat.
The single-copy Maxwell tensor $K_{\mu\nu}^+$ transforms in the same way as $F^+$ in \eqref{Ftransform}, and the transformation of its dual is the conjugate of this. It can be checked that the transformed tensors are (anti-)self-dual with respect to the transformed metric \eqref{newmetric}, and agree with the new Killing two form obtained from \eqref{Killingtwoform}  using the same Killing vector \eqref{EHKV} but with the index lowered with the new metric.
\\
Considering the Weyl tensor, we have the double copy relation for the Eguchi-Hanson metric
\begin{equation}
\label{WeylEH}
C^{EH}_{\mu\nu\rho\sigma} =  -\frac{uv-XY}{\lambda} \, C_{\mu\nu\rho\sigma} [K^+ ]   \,  ,
\end{equation}
where $K^+$ is the single copy Maxwell tensor \eqref{EHsc}. Note that the Weyl tensor for the Eguchi-Hanson metric is  self-dual. We find that the equivalents of  \eqref{EHErnsts} in the transformed metric are
\begin{equation}
\label{EHErnstsNew}
\sigma^{'+}= -\frac{2i\lambda}{(2\gamma\delta\lambda+i\delta^2(uv-XY))} \, ,\qquad \sigma^{'-} = \frac{1}{(4\gamma^2(uv-XY)+i\gamma\delta)} \, ,
\end{equation}
and that the double copy relationship is preserved by a general transformation, with
\begin{equation}
\label{WeylEHshiftedFull}
C^{'\pm}_{\mu\nu\rho\sigma} =  \frac{2}{\sigma^{'\pm}}\, C_{\mu\nu\rho\sigma} [K^{'\pm} ] \,  .
\end{equation}
\\
To gain some insight into the action of  $SL(2,\mathbb{R})$ in this example, consider the transformations with $\delta=0$. The shifted Maxwell fields in this case are given by
\begin{equation}
\label{KshiftedSD}
K^{'+}= \frac{1}{2  \lambda  \gamma^2} \Big( - du\wedge dv + \frac{(1- 8 \lambda \gamma^2)}{u} du\wedge d(XY) \Big) - 4 dX\wedge dY\,
\end{equation}
and
\begin{align}
\label{KshiftedASD}
K^{'-}=&  \frac{-uv+XY(1-16 \lambda\gamma^2)}{4(uv-XY)^3\gamma^2} du\wedge dv
 +\frac{Y(uv-XY + 8 \lambda \gamma^2(uv+XY))}{4 u (uv-XY)^3\gamma^2}du\wedge dX   \\     \nonumber
& +\frac{X(1- 8 \lambda \gamma^2)}{4u(uv-XY)^2\gamma^2} du \wedge dY
-\frac{4u \lambda X}{(uv-XY)^3} dv\wedge dY
+\frac{2\lambda(uv+XY)}{(uv-XY)^3} dX\wedge dY\, .
\end{align}
\\
For the transformed metric in this case, from  \eqref{WeylEHshiftedFull} the new Weyl tensor is now anti-self-dual and obeys the double copy relationship
$C^{'-}_{\mu\nu\rho\sigma} =  8 \gamma^2(uv-XY)\, C_{\mu\nu\rho\sigma} [K^{-'} ]
$, where $K^{-'}_{\mu\nu}$ is the shifted Maxwell tensor given in \eqref{KshiftedASD}. When the parameter $\gamma$ satisfies $8\lambda\gamma^2=1$ we find that $K_{\mu\nu}^{'-}= -K_{\mu\nu}^+$  (note that the dual of a tensor is defined with respect to different metrics on the two sides of this equation). Thus the roles of the electromagnetic fields $E$ and $B$ are exchanged by this transformation. The new metric is given simply in terms of the original Eguchi-Hanson metric by the interchange of $(u,v)\leftrightarrow(X,Y)$. The double copy relations are consistent with this. It would be interesting to follow up with a full study of the action of the transformations when $\delta\not= 0$.

%%%%%%%%%%%%%%%%%%%%%%%%%%%%%%%%%

\subsection{Higher dimensions}

Higher-dimensional double copies can be studied via a direct application of the formula \eqref{Dhigher}, but it should be kept in mind that in many cases of interest there is more structure in higher-dimensions  that may play a key role in consideration of double copies.
The review \cite{Frolov:2017kze}, for example, summarises work on symmetries of higher-dimensional Kerr-NUT-(A)dS black hole spacetimes.
These symmetries are linked with the presence of Killing and Killing-Yano tensors, which feature strongly in the discussion of the special properties of these spacetimes, such as their algebraic type, the  integrability of geodesic motion, and separability of the Hamilton-Jacobi, Klein-Gordon, and Dirac equations. The fundamental Killing object in these discussions is the ``principal tensor" $h$, which generates a complete set of explicit and hidden symmetries and uniquely determines the geometry of the Kerr-NUT-(A)dS metric.
\\

The principal tensor, and those other tensors generated by it, might be expected to also play a role in defining tensor double copies for such spacetimes in higher dimensions. If so, we would expect a relationship between any Maxwell tensor (or other two forms to be used in the double copy) and the principal tensor and its descendants. This might first be investigated in four dimensions. In the four-dimensional spacetime \eqref{typeDmetric} discussed  above, with the two null co-vectors
\begin{align}
K =  du + q^2 dv\,, \nonumber \\
L=  du - p^2 dv \,,
\end{align}
 the Maxwell single copy gauge field is given by
\begin{equation}
\label{SCMaxwell}
A= \frac{1}{p^2+q^2} \Big( g\, p K + e\, q L \Big)  \, .
\end{equation}
Its field strength is given by \eqref{MaxwellTN} above, {\it i.e.}
\begin{equation}
\label{MaxwellTNKilling}
F= \frac{1}{(p^2+q^2)^2} \Big( \alpha\, dp K + \beta\, dq L \Big)  \, ,
\end{equation}
with $\alpha = -2\,epq-e\,(p^2-q^2), \beta = -2\,gpq+e(p^2-q^2)$.
The principal tensor $h$ and its dual for this case are given by \cite{Frolov:2017kze}
\begin{align}
\label{KYTN}
h =  \frac{1}{(1-pq)^3} \Big( p\, dp K - q\, dq L \Big)\,, \nonumber \\
\tilde h =  \frac{1}{(1-pq)^3} \Big( q\, dp K +p\, dq L \Big)\, .
\end{align}
One can check that the Maxwell single copy \eqref{MaxwellTNKilling} can be expressed in terms of the principal tensor as %
\begin{equation}
\label{MaxwellKY}
F = a\, h + b\, \tilde h  \, ,
\end{equation}
where $a = \Omega^3(-3pq(gp+eq)+gq^3+ep^3)$ and $b = \Omega^3( - 3 p q (ep-gq) +eq^3-gp^3  )$, with $\Omega = \frac{1-pq}{p^2+q^2}$. Curiously, the factors in the coefficients $a, b$ (with $e\rightarrow m, g\rightarrow n$) also appear in some of the components of the Weyl tensor. The result  \eqref{MaxwellKY} implies that this four-dimensional Weyl double copy can be written purely in terms of the principal tensor. It would be interesting to explore higher-dimensional Weyl double copies by seeing how the principal tensor and its descendants, the related Killing tensors, along with the Maxwell tensor where this is independent, might be used in formul\ae\ like \eqref{Dhigher} to map to the Weyl tensor; spinor analogues might also be explored of course.

\section*{Discussion}
We have explored the double copy formalism in the context of solution generating symmetries in relativity, and in particular how these symmetries in general relativity are related to hidden or duality symmetries in the single copy gauge theory. We have used two complementary techniques, the Kerr-Schild and the Weyl double copy formalisms, and have seen in a number of cases that the transformations in the gauge theory and the corresponding transformations on the gravity side are such that  the double copy structure is preserved. This seems to us further evidence that the double copy is beyond just a perturbative symmetry for amplitudes but a fascinating relation between gravity and gauge theory.

\section*{Acknowledgments}
DSB and WJS are supported by the UK Science and Technology Facilities Council (STFC) with consolidated grant ST/L000415/1, {\it{String Theory, Gauge Theory and Duality}}. DPV is supported by a Royal Society studentship and RA is supported by a student scholarship from the Ministry of Higher Education of the UAE. We thank Joe Hayling, Ricardo Monteiro and Chris White for discussions on this topic.

\pagebreak

\bibliography{refs}

\providecommand{\href}[2]{#2}\begingroup\raggedright\begin{thebibliography}{10}

\bibitem{Bern:2008qj}
Z.~Bern, J.~Carrasco, and H.~Johansson, ``{New Relations for Gauge-Theory
  Amplitudes},'' {\em Phys.Rev.} {\bf D78} (2008) 085011,
\href{http://www.arXiv.org/abs/0805.3993}{{\tt 0805.3993}}.
%%CITATION = ARXIV:0805.3993;%%.

\bibitem{Bern:2010ue}
Z.~Bern, J.~J.~M. Carrasco, and H.~Johansson, ``{Perturbative Quantum Gravity
  as a Double Copy of Gauge Theory},'' {\em Phys.Rev.Lett.} {\bf 105} (2010)
  061602, \href{http://www.arXiv.org/abs/1004.0476}{{\tt 1004.0476}}.

\bibitem{Bern:2010yg}
Z.~Bern, T.~Dennen, Y.-t. Huang, and M.~Kiermaier, ``{Gravity as the Square of
  Gauge Theory},'' {\em Phys.Rev.} {\bf D82} (2010) 065003,
  \href{http://www.arXiv.org/abs/1004.0693}{{\tt 1004.0693}}.

\bibitem{BjerrumBohr:2009rd}
N.~Bjerrum-Bohr, P.~H. Damgaard, and P.~Vanhove, ``{Minimal Basis for Gauge
  Theory Amplitudes},'' {\em Phys.Rev.Lett.} {\bf 103} (2009) 161602,
\href{http://www.arXiv.org/abs/0907.1425}{{\tt 0907.1425}}.
%%CITATION = ARXIV:0907.1425;%%.

\bibitem{Stieberger:2009hq}
S.~Stieberger, ``{Open and Closed vs. Pure Open String Disk Amplitudes},''
\href{http://www.arXiv.org/abs/0907.2211}{{\tt 0907.2211}}.
%%CITATION = ARXIV:0907.2211;%%.

\bibitem{BjerrumBohr:2010zs}
N.~Bjerrum-Bohr, P.~H. Damgaard, T.~Sondergaard, and P.~Vanhove, ``{Monodromy
  and Jacobi-like Relations for Color-Ordered Amplitudes},'' {\em JHEP} {\bf
  1006} (2010) 003,
\href{http://www.arXiv.org/abs/1003.2403}{{\tt 1003.2403}}.
%%CITATION = ARXIV:1003.2403;%%.

\bibitem{Feng:2010my}
B.~Feng, R.~Huang, and Y.~Jia, ``{Gauge Amplitude Identities by On-shell
  Recursion Relation in S-matrix Program},'' {\em Phys.Lett.} {\bf B695} (2011)
  350--353,
\href{http://www.arXiv.org/abs/1004.3417}{{\tt 1004.3417}}.
%%CITATION = ARXIV:1004.3417;%%.

\bibitem{Tye:2010dd}
S.~Henry~Tye and Y.~Zhang, ``{Dual Identities inside the Gluon and the Graviton
  Scattering Amplitudes},'' {\em JHEP} {\bf 1006} (2010) 071,
\href{http://www.arXiv.org/abs/1003.1732}{{\tt 1003.1732}}.
%%CITATION = ARXIV:1003.1732;%%.

\bibitem{Mafra:2011kj}
C.~R. Mafra, O.~Schlotterer, and S.~Stieberger, ``{Explicit BCJ Numerators from
  Pure Spinors},'' {\em JHEP} {\bf 1107} (2011) 092,
\href{http://www.arXiv.org/abs/1104.5224}{{\tt 1104.5224}}.
%%CITATION = ARXIV:1104.5224;%%.

\bibitem{Monteiro:2011pc}
R.~Monteiro and D.~O'Connell, ``{The Kinematic Algebra From the Self-Dual
  Sector},'' {\em JHEP} {\bf 1107} (2011) 007,
\href{http://www.arXiv.org/abs/1105.2565}{{\tt 1105.2565}}.
%%CITATION = ARXIV:1105.2565;%%.

\bibitem{BjerrumBohr:2012mg}
N.~Bjerrum-Bohr, P.~H. Damgaard, R.~Monteiro, and D.~O'Connell, ``{Algebras for
  Amplitudes},'' {\em JHEP} {\bf 1206} (2012) 061,
\href{http://www.arXiv.org/abs/1203.0944}{{\tt 1203.0944}}.
%%CITATION = ARXIV:1203.0944;%%.

\bibitem{Kawai:1985xq}
H.~Kawai, D.~Lewellen, and S.~Tye, ``{A Relation Between Tree Amplitudes of
  Closed and Open Strings},'' {\em Nucl.Phys.} {\bf B269} (1986)
1.
%%CITATION = NUPHA,B269,1;%%.

\bibitem{Bern:1998ug}
Z.~Bern, L.~J. Dixon, D.~Dunbar, M.~Perelstein, and J.~Rozowsky, ``{On the
  relationship between Yang-Mills theory and gravity and its implication for
  ultraviolet divergences},'' {\em Nucl.Phys.} {\bf B530} (1998) 401--456,
\href{http://www.arXiv.org/abs/hep-th/9802162}{{\tt hep-th/9802162}}.
%%CITATION = HEP-TH/9802162;%%.

\bibitem{Green:1982sw}
M.~B. Green, J.~H. Schwarz, and L.~Brink, ``{N=4 Yang-Mills and N=8
  Supergravity as Limits of String Theories},'' {\em Nucl.Phys.} {\bf B198}
  (1982)
474--492.
%%CITATION = NUPHA,B198,474;%%.

\bibitem{Bern:1997nh}
Z.~Bern, J.~Rozowsky, and B.~Yan, ``{Two loop four gluon amplitudes in N=4
  superYang-Mills},'' {\em Phys.Lett.} {\bf B401} (1997) 273--282,
\href{http://www.arXiv.org/abs/hep-ph/9702424}{{\tt hep-ph/9702424}}.
%%CITATION = HEP-PH/9702424;%%.

\bibitem{Carrasco:2011mn}
J.~J. Carrasco and H.~Johansson, ``{Five-Point Amplitudes in N=4
  Super-Yang-Mills Theory and N=8 Supergravity},'' {\em Phys.Rev.} {\bf D85}
  (2012) 025006,
\href{http://www.arXiv.org/abs/1106.4711}{{\tt 1106.4711}}.
%%CITATION = ARXIV:1106.4711;%%.

\bibitem{Carrasco:2012ca}
J.~J.~M. Carrasco, M.~Chiodaroli, M.~Gunaydin, and R.~Roiban, ``{One-loop
  four-point amplitudes in pure and matter-coupled N=4 supergravity},'' {\em
  JHEP} {\bf 1303} (2013) 056,
\href{http://www.arXiv.org/abs/1212.1146}{{\tt 1212.1146}}.
%%CITATION = ARXIV:1212.1146;%%.

\bibitem{Mafra:2012kh}
C.~R. Mafra and O.~Schlotterer, ``{The Structure of n-Point One-Loop Open
  Superstring Amplitudes},'' {\em JHEP} {\bf 1408} (2014) 099,
\href{http://www.arXiv.org/abs/1203.6215}{{\tt 1203.6215}}.
%%CITATION = ARXIV:1203.6215;%%.

\bibitem{Boels:2013bi}
R.~H. Boels, R.~S. Isermann, R.~Monteiro, and D.~O'Connell,
  ``{Colour-Kinematics Duality for One-Loop Rational Amplitudes},'' {\em JHEP}
  {\bf 1304} (2013) 107,
\href{http://www.arXiv.org/abs/1301.4165}{{\tt 1301.4165}}.
%%CITATION = ARXIV:1301.4165;%%.

\bibitem{Bjerrum-Bohr:2013iza}
N.~E.~J. Bjerrum-Bohr, T.~Dennen, R.~Monteiro, and D.~O'Connell, ``{Integrand
  Oxidation and One-Loop Colour-Dual Numerators in N=4 Gauge Theory},'' {\em
  JHEP} {\bf 1307} (2013) 092,
\href{http://www.arXiv.org/abs/1303.2913}{{\tt 1303.2913}}.
%%CITATION = ARXIV:1303.2913;%%.

\bibitem{Bern:2013yya}
Z.~Bern, S.~Davies, T.~Dennen, Y.-t. Huang, and J.~Nohle, ``{Color-Kinematics
  Duality for Pure Yang-Mills and Gravity at One and Two Loops},''
\href{http://www.arXiv.org/abs/1303.6605}{{\tt 1303.6605}}.
%%CITATION = ARXIV:1303.6605;%%.

\bibitem{Bern:2013qca}
Z.~Bern, S.~Davies, and T.~Dennen, ``{The Ultraviolet Structure of Half-Maximal
  Supergravity with Matter Multiplets at Two and Three Loops},'' {\em
  Phys.Rev.} {\bf D88} (2013) 065007,
\href{http://www.arXiv.org/abs/1305.4876}{{\tt 1305.4876}}.
%%CITATION = ARXIV:1305.4876;%%.

\bibitem{Nohle:2013bfa}
J.~Nohle, ``{Color-Kinematics Duality in One-Loop Four-Gluon Amplitudes with
  Matter},''
\href{http://www.arXiv.org/abs/1309.7416}{{\tt 1309.7416}}.
%%CITATION = ARXIV:1309.7416;%%.

\bibitem{Bern:2013uka}
Z.~Bern, S.~Davies, T.~Dennen, A.~V. Smirnov, and V.~A. Smirnov, ``{Ultraviolet
  Properties of N=4 Supergravity at Four Loops},'' {\em Phys.Rev.Lett.} {\bf
  111} (2013), no.~23, 231302,
\href{http://www.arXiv.org/abs/1309.2498}{{\tt 1309.2498}}.
%%CITATION = ARXIV:1309.2498;%%.

\bibitem{Naculich:2013xa}
S.~G. Naculich, H.~Nastase, and H.~J. Schnitzer, ``{All-loop infrared-divergent
  behavior of most-subleading-color gauge-theory amplitudes},'' {\em JHEP} {\bf
  1304} (2013) 114,
\href{http://www.arXiv.org/abs/1301.2234}{{\tt 1301.2234}}.
%%CITATION = ARXIV:1301.2234;%%.

\bibitem{Du:2014uua}
Y.-J. Du, B.~Feng, and C.-H. Fu, ``{Dual-color decompositions at one-loop level
  in Yang-Mills theory},''
\href{http://www.arXiv.org/abs/1402.6805}{{\tt 1402.6805}}.
%%CITATION = ARXIV:1402.6805;%%.

\bibitem{Mafra:2014gja}
C.~R. Mafra and O.~Schlotterer, ``{Towards one-loop SYM amplitudes from the
  pure spinor BRST cohomology},'' {\em Fortsch.Phys.} {\bf 63} (2015), no.~2,
  105--131,
\href{http://www.arXiv.org/abs/1410.0668}{{\tt 1410.0668}}.
%%CITATION = ARXIV:1410.0668;%%.

\bibitem{Bern:2014sna}
Z.~Bern, S.~Davies, and T.~Dennen, ``{Enhanced Ultraviolet Cancellations in N =
  5 Supergravity at Four Loop},''
\href{http://www.arXiv.org/abs/1409.3089}{{\tt 1409.3089}}.
%%CITATION = ARXIV:1409.3089;%%.

\bibitem{Mafra:2015mja}
C.~R. Mafra and O.~Schlotterer, ``{Two-loop five-point amplitudes of super
  Yang-Mills and supergravity in pure spinor superspace},''
\href{http://www.arXiv.org/abs/1505.02746}{{\tt 1505.02746}}.
%%CITATION = ARXIV:1505.02746;%%.

\bibitem{He:2015wgf}
S.~He, R.~Monteiro, and O.~Schlotterer, ``{String-inspired BCJ numerators for
  one-loop MHV amplitudes},'' {\em JHEP} {\bf 01} (2016) 171,
\href{http://www.arXiv.org/abs/1507.06288}{{\tt 1507.06288}}.
%%CITATION = ARXIV:1507.06288;%%.

\bibitem{Bern:2015ooa}
Z.~Bern, S.~Davies, and J.~Nohle, ``{Double-Copy Constructions and Unitarity
  Cuts},''
\href{http://www.arXiv.org/abs/1510.03448}{{\tt 1510.03448}}.
%%CITATION = ARXIV:1510.03448;%%.

\bibitem{Mogull:2015adi}
G.~Mogull and D.~O'Connell, ``{Overcoming Obstacles to Colour-Kinematics
  Duality at Two Loops},'' {\em JHEP} {\bf 12} (2015) 135,
\href{http://www.arXiv.org/abs/1511.06652}{{\tt 1511.06652}}.
%%CITATION = ARXIV:1511.06652;%%.

\bibitem{Chiodaroli:2015rdg}
M.~Chiodaroli, M.~Gunaydin, H.~Johansson, and R.~Roiban, ``{Spontaneously
  Broken Yang-Mills-Einstein Supergravities as Double Copies},''
\href{http://www.arXiv.org/abs/1511.01740}{{\tt 1511.01740}}.
%%CITATION = ARXIV:1511.01740;%%.

\bibitem{Bern:2017ucb}
Z.~Bern, J.~J.~M. Carrasco, W.-M. Chen, H.~Johansson, R.~Roiban, and M.~Zeng,
  ``{The Five-Loop Four-Point Integrand of N=8 Supergravity as a Generalized
  Double Copy},''
\href{http://www.arXiv.org/abs/1708.06807}{{\tt 1708.06807}}.
%%CITATION = ARXIV:1708.06807;%%.

\bibitem{Johansson:2015oia}
H.~Johansson and A.~Ochirov, ``{Color-Kinematics Duality for QCD Amplitudes},''
  {\em JHEP} {\bf 01} (2016) 170,
\href{http://www.arXiv.org/abs/1507.00332}{{\tt 1507.00332}}.
%%CITATION = ARXIV:1507.00332;%%.

\bibitem{Oxburgh:2012zr}
S.~Oxburgh and C.~White, ``{BCJ duality and the double copy in the soft
  limit},'' {\em JHEP} {\bf 1302} (2013) 127,
\href{http://www.arXiv.org/abs/1210.1110}{{\tt 1210.1110}}.
%%CITATION = ARXIV:1210.1110;%%.

\bibitem{White:2011yy}
C.~D. White, ``{Factorization Properties of Soft Graviton Amplitudes},'' {\em
  JHEP} {\bf 1105} (2011) 060, \href{http://www.arXiv.org/abs/1103.2981}{{\tt
  1103.2981}}.

\bibitem{Melville:2013qca}
S.~Melville, S.~Naculich, H.~Schnitzer, and C.~White, ``{Wilson line approach
  to gravity in the high energy limit},'' {\em Phys.Rev.} {\bf D89} (2014)
  025009,
\href{http://www.arXiv.org/abs/1306.6019}{{\tt 1306.6019}}.
%%CITATION = ARXIV:1306.6019;%%.

\bibitem{Luna:2016idw}
A.~Luna, S.~Melville, S.~G. Naculich, and C.~D. White, ``{Next-to-soft
  corrections to high energy scattering in QCD and gravity},'' {\em JHEP} {\bf
  01} (2017) 052,
\href{http://www.arXiv.org/abs/1611.02172}{{\tt 1611.02172}}.
%%CITATION = ARXIV:1611.02172;%%.

\bibitem{Saotome:2012vy}
R.~Saotome and R.~Akhoury, ``{Relationship Between Gravity and Gauge Scattering
  in the High Energy Limit},'' {\em JHEP} {\bf 1301} (2013) 123,
\href{http://www.arXiv.org/abs/1210.8111}{{\tt 1210.8111}}.
%%CITATION = ARXIV:1210.8111;%%.

\bibitem{Vera:2012ds}
A.~Sabio~Vera, E.~Serna~Campillo, and M.~A. Vazquez-Mozo, ``{Color-Kinematics
  Duality and the Regge Limit of Inelastic Amplitudes},'' {\em JHEP} {\bf 1304}
  (2013) 086,
\href{http://www.arXiv.org/abs/1212.5103}{{\tt 1212.5103}}.
%%CITATION = ARXIV:1212.5103;%%.

\bibitem{Johansson:2013nsa}
H.~Johansson, A.~Sabio~Vera, E.~Serna~Campillo, and M.~. Vázquez-Mozo,
  ``{Color-Kinematics Duality in Multi-Regge Kinematics and Dimensional
  Reduction},'' {\em JHEP} {\bf 1310} (2013) 215,
\href{http://www.arXiv.org/abs/1307.3106}{{\tt 1307.3106}}.
%%CITATION = ARXIV:1307.3106;%%.

\bibitem{Johansson:2013aca}
H.~Johansson, A.~Sabio~Vera, E.~Serna~Campillo, and M.~A. Vazquez-Mozo,
  ``{Color-kinematics duality and dimensional reduction for graviton emission
  in Regge limit},''
\href{http://www.arXiv.org/abs/1310.1680}{{\tt 1310.1680}}.
%%CITATION = ARXIV:1310.1680;%%.

\bibitem{Monteiro:2014cda}
R.~Monteiro, D.~O'Connell, and C.~D. White, ``{Black holes and the double
  copy},'' {\em JHEP} {\bf 1412} (2014) 056,
\href{http://www.arXiv.org/abs/1410.0239}{{\tt 1410.0239}}.
%%CITATION = ARXIV:1410.0239;%%.

\bibitem{Luna:2015paa}
A.~Luna, R.~Monteiro, D.~O'Connell, and C.~D. White, ``{The classical double
  copy for Taub-NUT spacetime},'' {\em Phys. Lett.} {\bf B750} (2015) 272--277,
\href{http://www.arXiv.org/abs/1507.01869}{{\tt 1507.01869}}.
%%CITATION = ARXIV:1507.01869;%%.

\bibitem{Berman:2018hwd}
D.~S. Berman, E.~Chacon, A.~Luna, and C.~D. White, ``{The self-dual classical
  double copy, and the Eguchi-Hanson instanton},''
\href{http://www.arXiv.org/abs/1809.04063}{{\tt 1809.04063}}.
%%CITATION = ARXIV:1809.04063;%%.

\bibitem{Sabharwal:2019ngs}
S.~Sabharwal and J.~W. Dalhuisen, ``{Anti-Self-Dual Spacetimes, Gravitational
  Instantons and Knotted Zeros of the Weyl Tensor},'' {\em JHEP} {\bf 07}
  (2019) 004,
\href{http://www.arXiv.org/abs/1904.06030}{{\tt 1904.06030}}.
%%CITATION = ARXIV:1904.06030;%%.

\bibitem{Anastasiou:2014qba}
A.~Anastasiou, L.~Borsten, M.~J. Duff, L.~J. Hughes, and S.~Nagy, ``{Yang-Mills
  origin of gravitational symmetries},'' {\em Phys. Rev. Lett.} {\bf 113}
  (2014), no.~23, 231606,
\href{http://www.arXiv.org/abs/1408.4434}{{\tt 1408.4434}}.
%%CITATION = ARXIV:1408.4434;%%.

\bibitem{Borsten:2015pla}
L.~Borsten and M.~J. Duff, ``{Gravity as the square of Yang–Mills?},'' {\em
  Phys. Scripta} {\bf 90} (2015) 108012,
\href{http://www.arXiv.org/abs/1602.08267}{{\tt 1602.08267}}.
%%CITATION = ARXIV:1602.08267;%%.

\bibitem{Anastasiou:2016csv}
A.~Anastasiou, L.~Borsten, M.~J. Duff, M.~J. Hughes, A.~Marrani, S.~Nagy, and
  M.~Zoccali, ``{Twin supergravities from Yang-Mills theory squared},'' {\em
  Phys. Rev.} {\bf D96} (2017), no.~2, 026013,
\href{http://www.arXiv.org/abs/1610.07192}{{\tt 1610.07192}}.
%%CITATION = ARXIV:1610.07192;%%.

\bibitem{Anastasiou:2017nsz}
A.~Anastasiou, L.~Borsten, M.~J. Duff, A.~Marrani, S.~Nagy, and M.~Zoccali,
  ``{Are all supergravity theories Yang-Mills squared?},''
\href{http://www.arXiv.org/abs/1707.03234}{{\tt 1707.03234}}.
%%CITATION = ARXIV:1707.03234;%%.

\bibitem{Cardoso:2016ngt}
G.~L. Cardoso, S.~Nagy, and S.~Nampuri, ``{A double copy for $ \mathcal{N}=2 $
  supergravity: a linearised tale told on-shell},'' {\em JHEP} {\bf 10} (2016)
  127,
\href{http://www.arXiv.org/abs/1609.05022}{{\tt 1609.05022}}.
%%CITATION = ARXIV:1609.05022;%%.

\bibitem{Borsten:2017jpt}
L.~Borsten, ``{On $D=6$, $\mathcal{N}=(2,0)$ and $\mathcal{N}=(4,0)$
  theories},''
\href{http://www.arXiv.org/abs/1708.02573}{{\tt 1708.02573}}.
%%CITATION = ARXIV:1708.02573;%%.

\bibitem{Anastasiou:2017taf}
A.~Anastasiou, L.~Borsten, M.~J. Duff, A.~Marrani, S.~Nagy, and M.~Zoccali,
  ``{The Mile High Magic Pyramid},''
\newblock 2017.
\newblock
\href{http://www.arXiv.org/abs/1711.08476}{{\tt 1711.08476}}.
\newblock
%%CITATION = ARXIV:1711.08476;%%.

\bibitem{Anastasiou:2018rdx}
A.~Anastasiou, L.~Borsten, M.~J. Duff, S.~Nagy, and M.~Zoccali, ``{BRST
  squared},''
\href{http://www.arXiv.org/abs/1807.02486}{{\tt 1807.02486}}.
%%CITATION = ARXIV:1807.02486;%%.

\bibitem{LopesCardoso:2018xes}
G.~Lopes~Cardoso, G.~Inverso, S.~Nagy, and S.~Nampuri, ``{Comments on the
  double copy construction for gravitational theories},'' in {\em {17th
  Hellenic School and Workshops on Elementary Particle Physics and Gravity
  (CORFU2017) Corfu, Greece, September 2-28, 2017}}.
\newblock 2018.
\newblock
\href{http://www.arXiv.org/abs/1803.07670}{{\tt 1803.07670}}.
\newblock
%%CITATION = ARXIV:1803.07670;%%.

\bibitem{Montonen:1977sn}
C.~Montonen and D.~I. Olive, ``{Magnetic Monopoles as Gauge Particles?},'' {\em
  Phys. Lett.} {\bf 72B} (1977)
117--120.
%%CITATION = PHLTA,72B,117;%%.

\bibitem{PhysRev.115.1325}
H.~A. Buchdahl, ``Reciprocal static metrics and scalar fields in the general
  theory of relativity,'' {\em Phys. Rev.} {\bf 115} (Sep, 1959) 1325--1328.

\bibitem{Luna:2018dpt}
A.~Luna, R.~Monteiro, I.~Nicholson, and D.~O'Connell, ``{Type D Spacetimes and
  the Weyl Double Copy},'' {\em Class. Quant. Grav.} {\bf 36} (2019) 065003,
\href{http://www.arXiv.org/abs/1810.08183}{{\tt 1810.08183}}.
%%CITATION = ARXIV:1810.08183;%%.

\bibitem{Huang:2019cja}
Y.-t. Huang, U.~Kol, and D.~O'Connell, ``{The Double Copy of Electric-Magnetic
  Duality},''
\href{http://www.arXiv.org/abs/1911.06318}{{\tt 1911.06318}}.
%%CITATION = ARXIV:1911.06318;%%.

\bibitem{Taub:1950ez}
A.~H. Taub, ``{Empty space-times admitting a three parameter group of
  motions},'' {\em Annals Math.} {\bf 53} (1951)
472--490.
%%CITATION = ANMAA,53,472;%%.

\bibitem{Newman:1963yy}
E.~Newman, L.~Tamburino, and T.~Unti, ``{Empty space generalization of the
  Schwarzschild metric},'' {\em J. Math. Phys.} {\bf 4} (1963)
915.
%%CITATION = JMAPA,4,915;%%.

\bibitem{Ortin:2015hya}
T.~Ortin, {\em {Gravity and Strings}}.
\newblock Cambridge Monographs on Mathematical Physics. Cambridge University
  Press,
2015.
\newblock
%%CITATION = INSPIRE-1383727;%%.

\bibitem{PLEBANSKI1975196}
J.~F. Plebañski, ``A class of solutions of einstein-maxwell equations,'' {\em
  Annals of Physics} {\bf 90} (1975), no.~1, 196 -- 255.

\bibitem{CHONG2005124}
Z.-W. Chong, G.~Gibbons, H.~Lü, and C.~Pope, ``Separability and killing
  tensors in kerr–taub-nut–de sitter metrics in higher dimensions,'' {\em
  Physics Letters B} {\bf 609} (2005), no.~1, 124 -- 132.

\bibitem{Ernst:1967wx}
F.~J. Ernst, ``{New formulation of the axially symmetric gravitational field
  problem},'' {\em Phys. Rev.} {\bf 167} (1968)
1175--1179.
%%CITATION = PHRVA,167,1175;%%.

\bibitem{Ehlers:1961zza}
J.~Ehlers, ``{Transformations of static exterior solutions of Einstein's
  gravitational field equations into different solutions by means of conformal
  mapping},'' {\em Colloq. Int. CNRS} {\bf 91} (1962)
275--284.
%%CITATION = COINA,91,275;%%.

\bibitem{Momeni:2005uc}
D.~Momeni, M.~Nouri-Zonoz, and R.~Ramezani-Arani, ``{MM-NUT disk space via
  Ehlers transformation},'' {\em Phys. Rev.} {\bf D72} (2005) 064023,
\href{http://www.arXiv.org/abs/gr-qc/0508036}{{\tt gr-qc/0508036}}.
%%CITATION = GR-QC/0508036;%%.

\bibitem{Geroch}
R.~Geroch, ``A method for generating solutions of einstein's equations,'' {\em
  Journal of Mathematical Physics} {\bf 12} (1971), no.~6, 918--924.

\bibitem{Monteiro:2018xev}
R.~Monteiro, I.~Nicholson, and D.~O'Connell, ``{Spinor-helicity and the
  algebraic classification of higher-dimensional spacetimes},''
\href{http://www.arXiv.org/abs/1809.03906}{{\tt 1809.03906}}.
%%CITATION = ARXIV:1809.03906;%%.

\bibitem{Plebanski:1976gy}
J.~F. Plebanski and M.~Demianski, ``{Rotating, charged, and uniformly
  accelerating mass in general relativity},'' {\em Annals Phys.} {\bf 98}
  (1976)
98--127.
%%CITATION = APNYA,98,98;%%.

\bibitem{Griffiths:2005qp}
J.~B. Griffiths and J.~Podolsky, ``{A New look at the Plebanski-Demianski
  family of solutions},'' {\em Int. J. Mod. Phys.} {\bf D15} (2006) 335--370,
\href{http://www.arXiv.org/abs/gr-qc/0511091}{{\tt gr-qc/0511091}}.
%%CITATION = GR-QC/0511091;%%.

\bibitem{Mars:2001gd}
M.~Mars, ``{Space-time Ehlers group: Transformation law for the Weyl tensor},''
  {\em Class. Quant. Grav.} {\bf 18} (2001) 719--738,
\href{http://www.arXiv.org/abs/gr-qc/0101020}{{\tt gr-qc/0101020}}.
%%CITATION = GR-QC/0101020;%%.

\bibitem{Mars:2016ynw}
M.~Mars, T.-T. Paetz, and J.~M.~M. Senovilla, ``{Classification of Kerr–de
  Sitter-like spacetimes with conformally flat ${I}$},'' {\em Class. Quant.
  Grav.} {\bf 34} (2017), no.~9, 095010,
\href{http://www.arXiv.org/abs/1610.09846}{{\tt 1610.09846}}.
%%CITATION = ARXIV:1610.09846;%%.

\bibitem{Gomez-Lobo:2016ykv}
A.~G.-P. Gomez-Lobo, ``{Vacuum type D initial data},'' {\em Class. Quant.
  Grav.} {\bf 33} (2016), no.~17, 175005,
  \href{http://www.arXiv.org/abs/1602.08075}{{\tt 1602.08075}}.
[Erratum: Class. Quant. Grav.35,no.7,079501(2018)].
%%CITATION = ARXIV:1602.08075;%%.

\bibitem{Paetz:2017lkn}
T.-T. Paetz, ``{Algorithmic characterization results for the Kerr-NUT-(A)dS
  space-time. I. A space-time approach},'' {\em J. Math. Phys.} {\bf 58}
  (2017), no.~4, 042501,
\href{http://www.arXiv.org/abs/1701.02959}{{\tt 1701.02959}}.
%%CITATION = ARXIV:1701.02959;%%.

\bibitem{Paetz:2017cai}
T.-T. Paetz, ``{Algorithmic characterization results for the Kerr-NUT-(A)dS
  space-time. II. KIDs for the Kerr-(A)(de Sitter) family},'' {\em J. Math.
  Phys.} {\bf 58} (2017), no.~4, 042502,
\href{http://www.arXiv.org/abs/1701.03315}{{\tt 1701.03315}}.
%%CITATION = ARXIV:1701.03315;%%.

\bibitem{Frolov:2017kze}
V.~Frolov, P.~Krtous, and D.~Kubiznak, ``{Black holes, hidden symmetries, and
  complete integrability},'' {\em Living Rev. Rel.} {\bf 20} (2017), no.~1, 6,
\href{http://www.arXiv.org/abs/1705.05482}{{\tt 1705.05482}}.
%%CITATION = ARXIV:1705.05482;%%.

\end{thebibliography}\endgroup

\end{document}